\documentclass[usenatbib]{mnras}
\pdfoutput=1

\usepackage{newtxtext,newtxmath}

\usepackage[T1]{fontenc}

\DeclareRobustCommand{\VAN}[3]{#2}
\let\VANthebibliography\thebibliography
\def\thebibliography{\DeclareRobustCommand{\VAN}[3]{##3}\VANthebibliography}

\usepackage{graphicx}	
\usepackage{amsmath}	
\usepackage{amssymb}	
\usepackage{float}
\usepackage{subfig}
\usepackage{enumitem}
\usepackage{booktabs}
\usepackage{xcolor}
\usepackage{notoccite}
\newlength\mylen
\setlist[itemize,1]{leftmargin=*,labelwidth=*,labelsep=-\mylen}

\newcommand{\msol}{M$_{\odot}$}
\newcommand{\kms}{km\,s$^{-1}$}

\title{HI study of isolated and pair galaxies: the MIR SFR-M$\star$ sequence}

\author[Bok et al.]{
J. Bok,$^{1,2}$\thanks{E-mail: jamie@ast.uct.ac.za}
R.E. Skelton,$^{1}$
M.E. Cluver,$^{3,4}$
T.H. Jarrett, $^{2}$
M.G. Jones, $^{5}$ 
\newauthor
 \hspace{1mm}and L. Verdes-Montenegro $^{5}$
\\
$^{1}$South African Astronomical Observatory, Observatory, Cape Town, 7935,  South Africa \\
$^{2}$University of Cape Town, Rondebosch, Cape Town, 7700, South Africa \\
$^{3}$Centre for Astrophysics and Supercomputing, Swinburne University of Technology, Hawthorn, Victoria 3122, Australia \\
$^{4}$Department of Physics and Astronomy, University of the Western Cape,Robert Sobukwe Road, Bellville, South Africa\\
$^{5}$Instituto de Astrof\'isica de Andaluc\'{i}a (IAA-CSIC), Glorieta de la Astronom\'{i}a, 18008 Granada, Spain}

\date{Accepted XXX. Received YYY; in original form ZZZ}

\pubyear{2015}

\begin{document}

\label{firstpage}
\pagerange{\pageref{firstpage}--\pageref{lastpage}}
\maketitle

\begin{abstract}
Using mid-infrared star formation rate and stellar mass indicators in $\textit{WISE}$, we construct and contrast the relation between star formation rate and stellar mass for isolated and paired galaxies. Our samples comprise a selection of AMIGA (isolated galaxies) and pairs of ALFALFA galaxies with HI detections such that we can examine the relationship between HI content (gas fraction, HI deficiency) and galaxy location on the main sequence (MS) in these two contrasting environments. We derive for the first time an HI scaling relation for isolated galaxies using $\textit{WISE}$ stellar masses, and thereby establish a baseline predictor of HI content that can be used to assess the impact of environment on HI content when compared with samples of galaxies in different environments. We use this updated relation to determine the HI deficiency of both our paired and isolated galaxies. Across all the quantities examined as a function of environment in this work (MS location, gas fraction, and HI deficiency), the AMIGA sample of isolated galaxies is found to have the lower dispersion: $\sigma_{\rm{AMIGA}} = 0.37$ versus $\sigma_{\rm{PAIRS}} = 0.55$ on the MS, $\sigma_{\rm{AMIGA}} = 0.44$ versus $\sigma_{\rm{PAIRS}} = 0.54$ in gas fraction, and $\sigma_{\rm{AMIGA}} = 0.28$ versus $\sigma_{\rm{PAIRS}} = 0.34$ in HI deficiency. We also note fewer isolated quiescent galaxies, 3 (0.6$\%$), compared to 12 (2.3$\%$) quiescent pair members. Our results suggest the differences in scatter measured between our samples are environment driven. Galaxies in isolation behave relatively predictably, and galaxies in more densely populated environments adopt a more stochastic behaviour, across a broad range of quantities.

\end{abstract}


\section{Introduction}
The correlation between a galaxy's star formation rate (SFR) and its stellar mass (M$\star$) (the `galaxy main sequence' or MS) is well established, with numerous studies in the past decade demonstrating it to be both real, and able to provide keen insight into the processes driving star formation (SF) in galaxies (\cite{Noeske2007,Bouche2010,Lee2015}).  The galaxy MS diagram shows a trend of increasing SFR with increasing M$\star$ that begins to plateau for M$\star> 10^{10.5}$ \msol$~$  \citep{Noeske2007}. This value for stellar mass is where the break in the stellar mass-halo mass relation occurs via virial shock heating \citep{Behroozi2013}.
 As we look to higher redshifts, the shape of this relationship remains similar, however with the entire sequence evolving to higher SFRs \citep{Lee2015, Whitaker2015, Speagle2014}. The bulk of SF thus appears to have occurred earlier in massive galaxies compared to less massive systems \citep{Noeske2007}. For a spiral galaxy, any deviations from the main sequence likely indicate a break from secular evolution. Quenching events can cause a galaxy to drop suddenly below the MS, while a starburst phase will cause a galaxy to migrate upwards off the MS.\\ 
Galaxy structure is also correlated with SFR. Numerous observations have revealed star-forming galaxies to be less concentrated than quiescent galaxies at a fixed mass \citep{Whitaker2015}. The causal relationship between a galaxy's morphology and its SF history, however, is not well understood. The process by which a blue star-forming spiral galaxy transforms into a red and dead elliptical galaxy, the physical mechanism that quenches SF in galaxies, remains one of the biggest puzzle pieces still missing from our picture of galaxy evolution. Current theories for what shuts down SF in galaxies can be categorized into two main classes, internal processes (starburst and AGN feedback \citep{McNamara2000, McNamara&Nulsen2007, Cicone2014}, mass quenching, and ``morphological quenching'' \citep{Martig2009}) and external processes, e.g. harassment \citep{Moore1996}, ram-pressure stripping \citep{Gunn1972}, viscous stripping \citep{Nulsen1982}, and strangulation \citep{Peng2015}. In contrast to quenching, starburst galaxies are also observed. During starburst events galaxies undergo rapid SF causing them to jump upwards off the MS \cite{Rodighiero2011,Elbaz2011,Schreiber2015}. These events are short lived, and galaxies likely ultimately settle back onto the MS, however with substantial additional mass. The circumstances leading to such events remain unclear, as well as the prevalence of this phenomenon.

\cite{McPartland2018}, in their investigation into quenching via internal processes, found the SFR-M$\star$ relation to be linear for a sample of star-forming galaxies from the DR7 SDSS catalogue with $H\alpha$-derived SFRs. Inclusion of Composite, Seyfert 2, and LINER galaxies on the MS, however, produces a flattening in the relation at M$\star>10^{10.5}~$M$\odot$, beyond which a negative slope is observed. Furthermore it was found that the reddest galaxies with the largest B/T ratios had SFRs $> 1.5$ dex below the MS, leading \cite{McPartland2018} to conclude that the quenching of SF in massive galaxies is strongly associated with AGN activity, with bulge growth an important pathway to this quenching.  Similar results are reported by \cite{Morselli2018} who studied the spatial distribution of both stellar mass and SF activity in a sample of 712 galaxies at 0.2<z<1.2 (excluding AGN and merging systems) from the GOODS field \citep{Giavalisco2004} with high resolution UV data from the Hubble Deep UV Survey. Galaxies above the MS were found to be more extended compared to their MS counterparts at fixed stellar masses, while galaxies below the MS, where SF was centrally suppressed, were mostly bulge-dominated. Quiescent galaxies, defined as those lying 1 dex below the MS, exhibited the highest instances of central SF suppression. With SF centrally enhanced and suppressed in galaxies above and below the MS respectively, the results of \cite{Morselli2018} support the scenario in which galaxies quench from the inside out. 

\cite{Wang2018} also investigated the link between galaxy structure and quenching, and incorporate the role of HI in their study. Using a sample of $\sim$1600 galaxies from the NASA-Sloan-Atlas catalogue within the ALFALFA footprint, \cite{Wang2018} investigated the two-step quenching scenario of compaction and quenching by comparing SFRs of compact star-forming galaxies (cSFGs) with extended star-forming galaxies (eSFGs). It was found that at fixed stellar mass the cSFGs exhibited similar or slightly higher SFRs compared to the eSFGs, as well as higher gas-phase metallicities. Moreover, the eSFGs were found on average to be more gas rich than the cSFGs, and had a median HI gas-depletion timescale of $\sim$8 Gyr compared to a median gas depletion of time of $\sim$4 Gyr for the cSFGs. With the environments of the eSFGs and cSFGs indistinguishable in their sample, \cite{Wang2018} conclude that galaxies evolve from eSFGs to cSFGs before joining the quenched population via an environment independent scenario of compaction and quenching.  \cite{Bitsakis2019} looked at the MS for galaxies from the CALIFA survey and found the cessation of SF to be the combined result of gas deficiency and the inefficiency of the remaining gas to form new stars. The authors attribute the latter to the build-up of a bulge component (morphological evolution). In their exploration of alternative SF suppression mechanisms, \cite{Bitsakis2019} found that the action of bars, AGN activity, local galaxy environment, as well as galaxy mergers, have only temporary effects on current SF, and are not responsible for the permanent quenching of SF. \\
Studying SF as a function of redshift, \cite{Tacconi2013} find molecular gas content principally responsible for the cosmic evolution of the SFR. Using CO measurements provided by the PHIBSS survey and looking at the MS from the perspective of molecular gas content, \cite{Tacconi2018} view gas fraction as a measure of a galaxys' gas accretion rate  (regulated by mergers/fluctuations in gas transport along the cosmic web), and SF efficiency as related to internal galaxy properties. Their study shows SF efficiency to remain roughly constant as stellar mass increases, while gas fractions decrease. These results again point to the importance of considering the gas components of galaxies when seeking to understand their location above, below, or on the MS.

The above mentioned studies have thus far either neglected galaxy environment, or found it to be unimportant with respect to the quenching of SF. \cite{Cochrane2018}, however, find strong evidence for environment-driven quenching in satellite galaxies. Probing the roles of mass and environment quenching in galaxy evolution, \cite{Cochrane2018} studied the relationship between halo mass, stellar mass, and SFR using the pioneering cosmological hydrodynamical simulation, EAGLE. \cite{Ellison2010} also look at the role of environment in enhancing/suppressing SF by studying galaxy pairs in SDSS. Instances of triggered SF were only observed in low-to-medium density environments, which the authors attribute to the higher gas fractions typical of low density environments, while mergers in high density environments are mainly without SF. The recent study of \cite{Pearson2019} compared SFRs between large samples of merging and non-merging galaxy systems from the KiDS and CANDELS imaging surveys. While large differences in SFR were occasionally observed, galaxy mergers in general were found to have little impact on SFRs, causing only minor shifts of $\sim$ 0.1 dex above and below the MS. \cite{Moon2019} further investigate the influence of companions on SFR by considering different types of companions, namely SF companions versus quiescent. Their study shows quiescent neighbours acting to suppress SF in their counterparts, while SF neighbours enhance SF in their companions. Both effects are enhanced as the projected separation between companions is reduced. An earlier study by \cite{Xu2010} studied SFR enhancement in close galaxy pairs in the Local Universe, however distinguished between spiral pairs (S+S) and spiral galaxies paired with ellipticals (S+E). No enhancement in SFR was observed in the S+E pairs, and the SFR enhancement observed in the S+S pairs was shown to be highly mass dependent, with significant SFR enhancement only occurring in massive (M$_{\star} > 10^{10}$) S+S pairs. A follow up study by \cite{Xu2012} revealed a negative cosmic evolution of SFR enhancement in S+S pairs, and attribute this to the trend of increasing gas fraction with redshift. In high gas fraction scenarios the gas disk experiences less gravitational torque from the stellar disk, and less disk gas is funnelled to the nucleus via loss of angular momentum. 

While the literature reports conflicting conclusions as to the role of environment in determining a galaxy's location on the MS (above or below, enhanced/suppressed SF), the location of isolated galaxies on the MS has yet to be determined. Constructing a sample of truly isolated galaxies is a non-trivial endeavour, one to which the AMIGA project (Analysis of the interstellar Medium in Isolated GAlaxies \cite{VerdesMontenegro2005}) is dedicated. The AMIGA project is an in depth multiwavelength study of isolated galaxies in order to distinguish galaxy properties arising from secular evolution, from those which result from external influences (see Section 2.1 for details). It can thus be used to probe how galaxy interactions might enhance/suppress SF, which we ascertain by their location on the MS diagram relative to the baseline AMIGA MS.

A key objective of this work is to construct a MS for isolated galaxies that can be used as a reference for different galaxy samples to probe the role of environment in governing galaxy location on the MS. In this way we also seek to gain insight into the processes responsible for the shutting down of SF in a nurture-free environment, where one is unable to point the finger at an interfering companion. Since interacting systems are inherently dusty by nature, construction of our MS makes use of MIR indicators in \textit{WISE} (W3 is sensitive to warm dust) to compute stellar masses and SFRs. Details of these measurements can be found in Section 3.

This paper is organised as follows: Our sample selection and data is described in Section 2, and the updated AMIGA HI scaling relation is presented in Section 3. We put forward and discuss our results in Section 4, namely the MS for our isolated and paired galaxies in Section 4.1, gas fraction in Section 4.2, and HI deficiency in Section 4.3. In Section 5 we discuss the relationship between gas (gas fraction and HI deficiency) and quenching in galaxies. Our main results and conclusions are summarised in Section 6. Throughout this paper we adopt a $\Lambda CDM$ cosmology with $H_0=70 $ \kms Mpc$^{-1}$, $\Omega_{M}= 0.3$, and $\Omega_{\Lambda} = 0.7$.

\section{Sample}
\subsection{Sample of isolated galaxies}
Our sample of isolated galaxies comprises galaxies selected from the AMIGA project. The AMIGA project provides the most extensive multiwavelength study of a well defined sample of isolated galaxies drawn from the \cite{Karachentseva1973}'s Catalogue of Isolated Galaxies (CIG). The original criteria of \cite{Karachentseva1973} classify a galaxy as isolated only if it is separated (in projection) from any neighbouring galaxy of isophotal diameter of between 1/4 and 4 times its own diameter (in B-band) by at least 20 times the diameter of the potential neighbour, that is, it is separated from the largest of these potential neighbours by at least 80 times its own diameter.
As part of the AMIGA project the degree of isolation was first re-evaluated by \cite{Verley2007a} by analysing Digitised POSS-I E images of each CIG galaxy and constructing a catalogue of potential neighbours. 
Quantification of the degree of isolation was introduced by \cite{Verley2007} by way of two complimentary isolation parameters, namely $\eta$ and Q, which measure the projected surface density of neighbours out to the 5th nearest neighbour and the tidal force exerted by neighbouring galaxies, respectively.
Cuts based on these two parameters were enforced to eliminate any outlying cases which may not be strictly isolated. In addition any CIG with a heliocentric redshift of less than 1500 km/s was removed as isolation is difficult to ensure for such nearby targets as a very large sky area must be considered. 
This left $\sim$700 galaxies in the AMIGA sample that adhere to strict isolation criteria such that for the past $\sim$3 Gyr they are unlikely to have interacted with any neighbour of significant mass \citep{VerdesMontenegro2005}.
\cite{Argudo-Fernandez2013} reassessed the isolation of AMIGA galaxies using both photometric and spectroscopic data from the ninth data release of the Sloan Digital Sky Survey (SDSS-DR9). The $\eta$ and Q metrics were recalculated based on the SDSS images, which led to an additional 16 per cent of the CIG galaxies failing the isolation criteria due to fainter neighbours being identified in the SDSS images. However, in the fields with spectroscopic data the same metrics were recalculated, but with the requirement that any neighbours must be within 500 km/s in redshift, which resulted in an equivalent fraction of the CIG galaxies being classified as isolated as in \cite{Verley2007a}. Furthermore, galaxies that are classified as isolated by one of these works are not necessarily classified as isolated in the other. Therefore, given that the \cite{Verley2007} analysis covers the full sample, whereas \cite{Argudo-Fernandez2013} misses a significant fraction due to the limits of the SDSS footprint, we draw our sample of isolated galaxies from the $\sim$700 galaxies that meet the \cite{Verley2007} criteria. The AMIGA team have further characterized the different components/phases of the interstellar medium (ISM) of these isolated galaxies, as well as their stellar components, in various different wavelengths, including a) optical \citep{VerdesMontenegro2005}, b) FIR \citep{Lisenfeld2007}, c) radio-continuum \citep{Leon2008}, and d) H$\alpha$ emission \citep{Verley2007a}, as well as e) nuclear activity \citep{Sabater2008, Sabater2011}. (Full details of the AMIGA project and their extensive research on isolated galaxies are available at http://amiga.iaa.es)

HI data is available for a sub-sample of the full AMIGA catalogue, the AMIGA HI science sample \citep{Jones2018}. We restrict our study to the HI science sample to assess the role of HI in determining the MS locations of our isolated galaxies. The AMIGA HI science sample \citep{Jones2018} is an HI subset of \cite{Karachentseva1973}'s 1050 CIG galaxy catalogue. HI spectra from the literature were compiled for 415 of these galaxies (see Table 1 in \cite{Jones2018} for observation details), and AMIGA conducted their own observations of 488 galaxies using the Arecibo, Effelsberg, Greenbank, and Nan\c{c}ay radio telescopes. (A summary of the AMIGA HI observations is displayed in Table 2 of \cite{Jones2018}.) 429 of the 488 CIG galaxies observed by AMIGA made it into the HI catalogue, and together with the 415 galaxies with spectra published in the literature, the final AMIGA HI science sample comprises 844 galaxies in total. All spectral parameters of this catalogue were extracted using the same fitting method, including the cases where existing observations from the literature were used, and as such this catalogue is considered a highly uniform HI database of isolated galaxies. \cite{Jones2018} further provide cuts to the HI sample based on completeness, isolation, and profile quality. For the purposes of our study we select only those galaxies within the complete AMIGA HI science sample that have been flagged as reliably isolated, with high quality HI profiles (544 galaxies in total). Morphological classification of the AMIGA sample conducted by \cite{Sulentic2006}, and later revisions by \cite{Buta2019}, show the AMIGA sample to be spiral galaxy dominated, specifically intermediate to late type (Sb-Sc). A photometric analysis conducted by \cite{Durbala2008} revealed the sample to be more symmetric, less concentrated, and less clumpy when compared to galaxies selected without isolated criteria. 

\begin{figure*}
\begin{tabular}{|c|c|}
\includegraphics[width = 38mm]{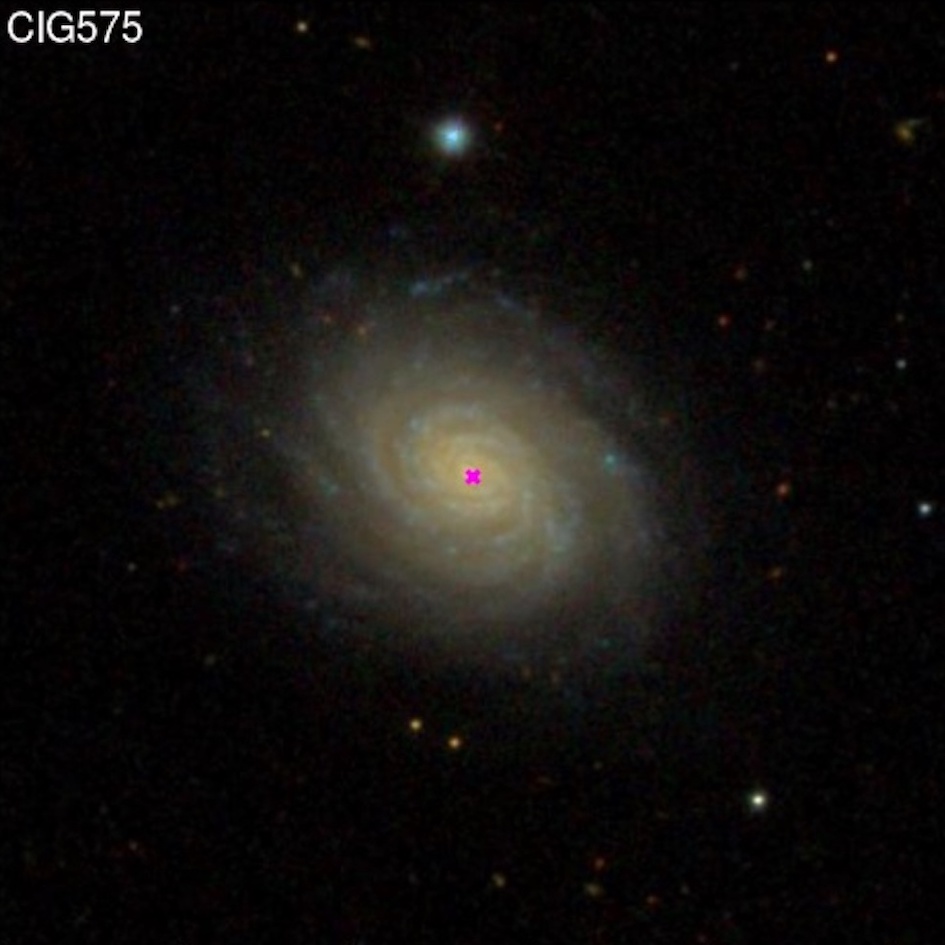} &  \includegraphics[width = 38mm]{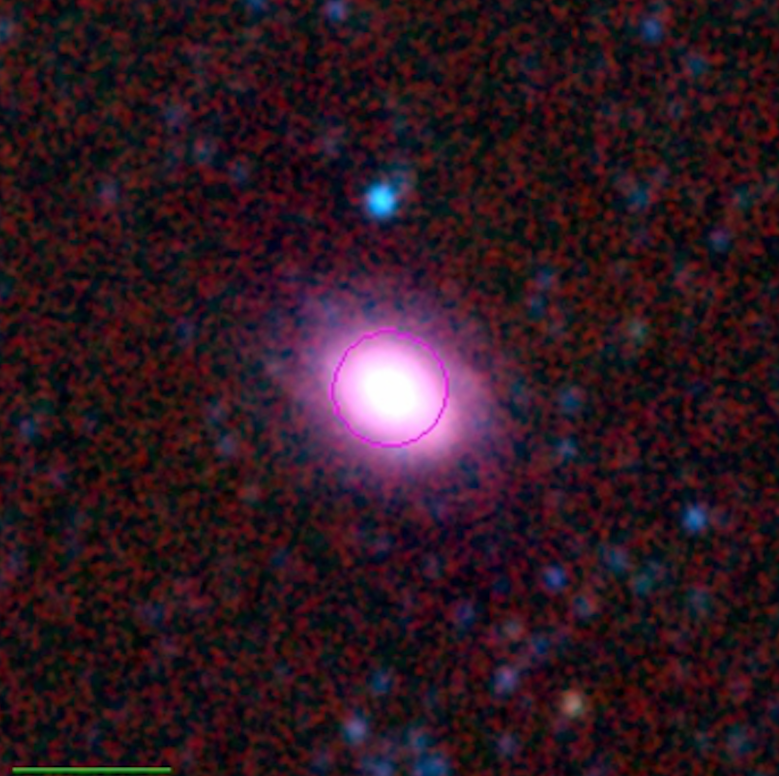} \\
\includegraphics[width = 38mm]{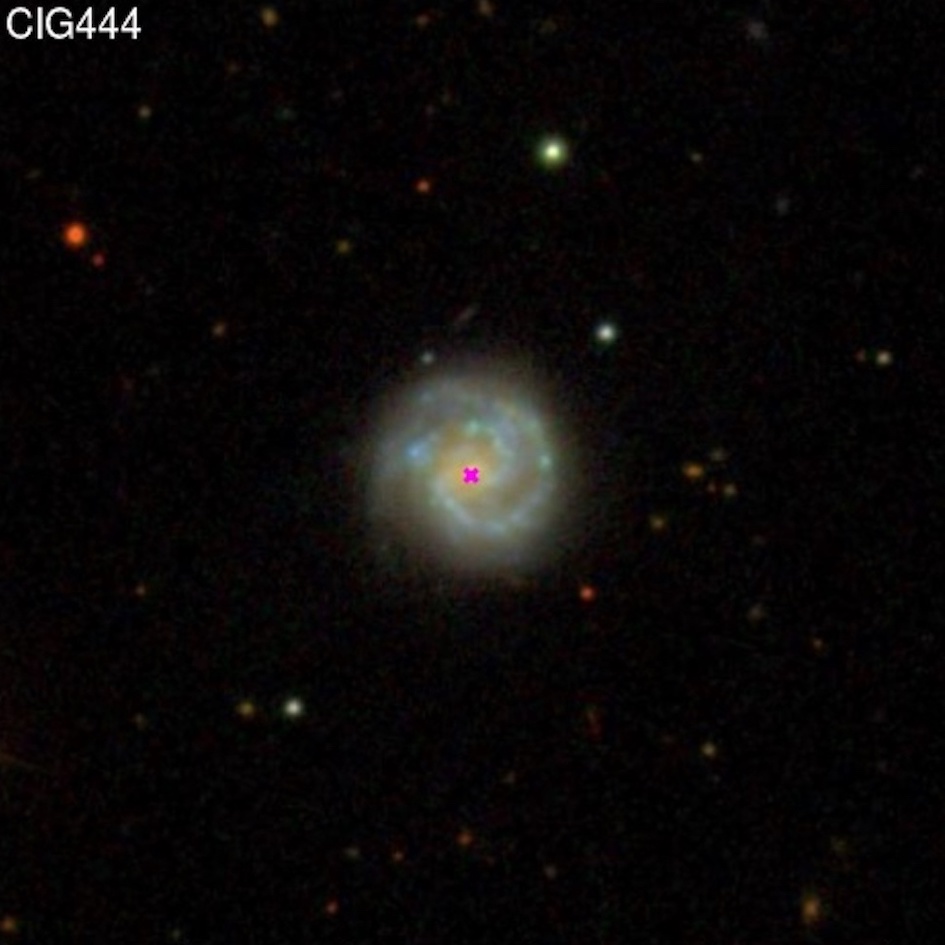} &     \includegraphics[width = 38mm]{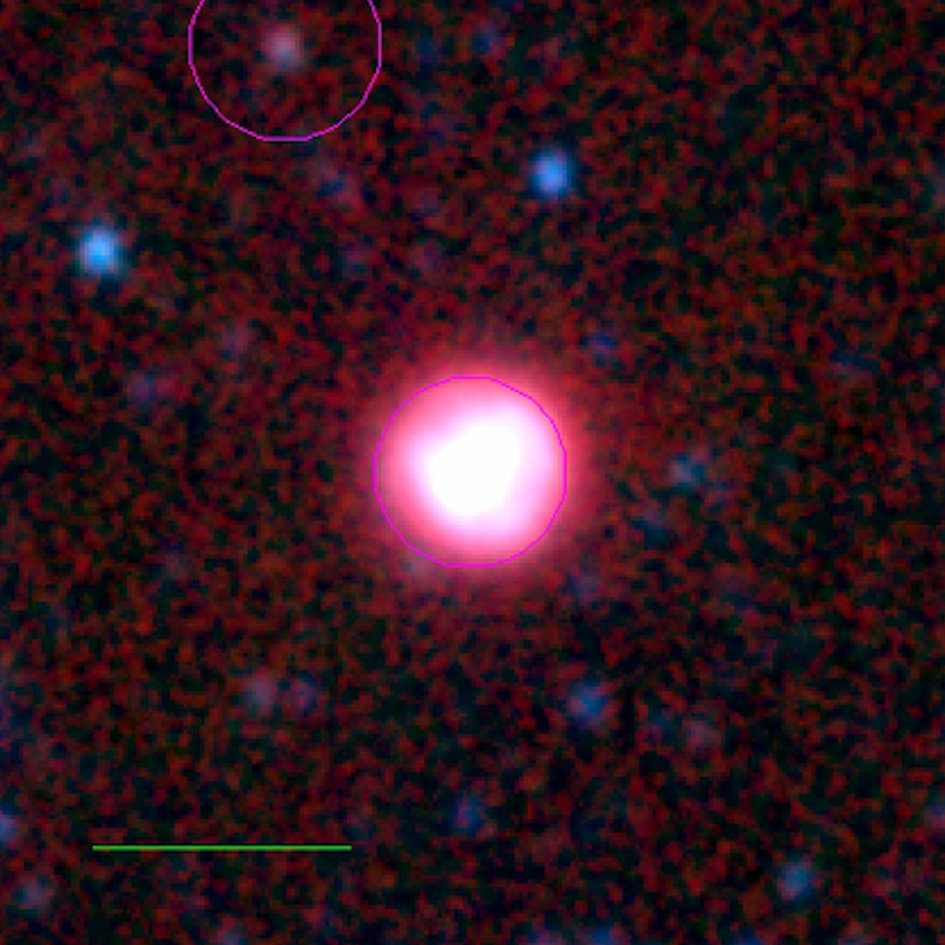} \\
 \includegraphics[width = 38mm]{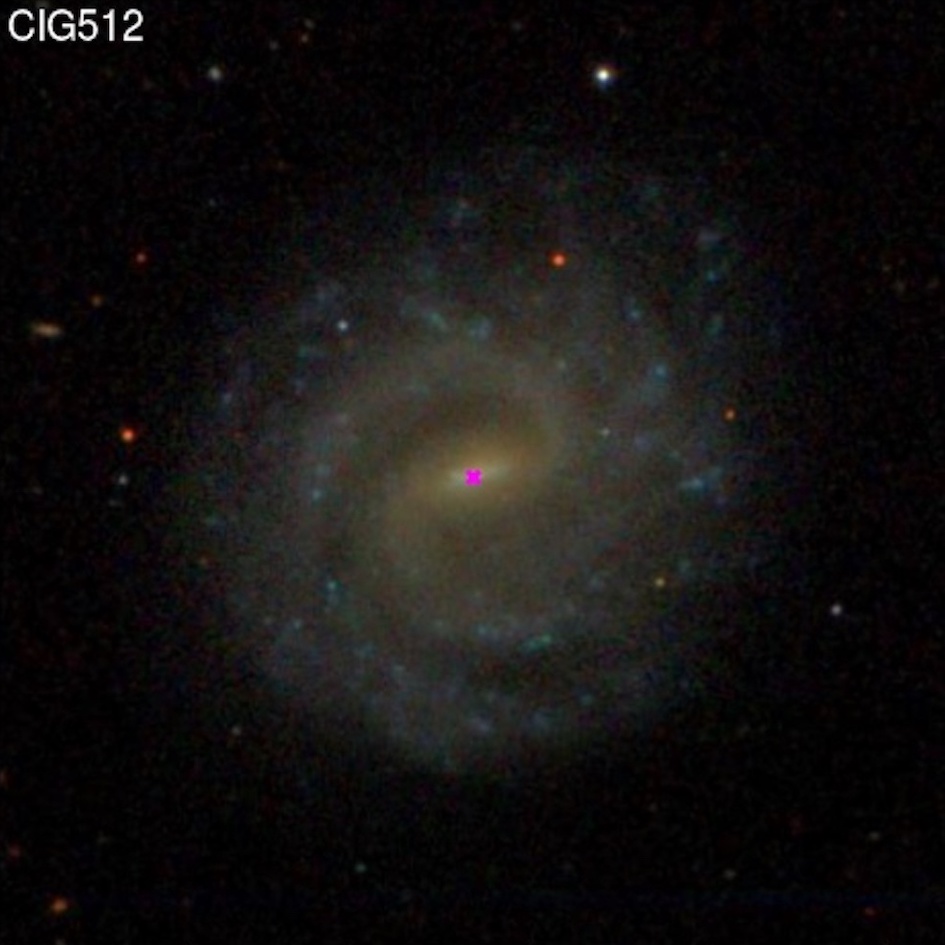}& \includegraphics[width = 38mm]{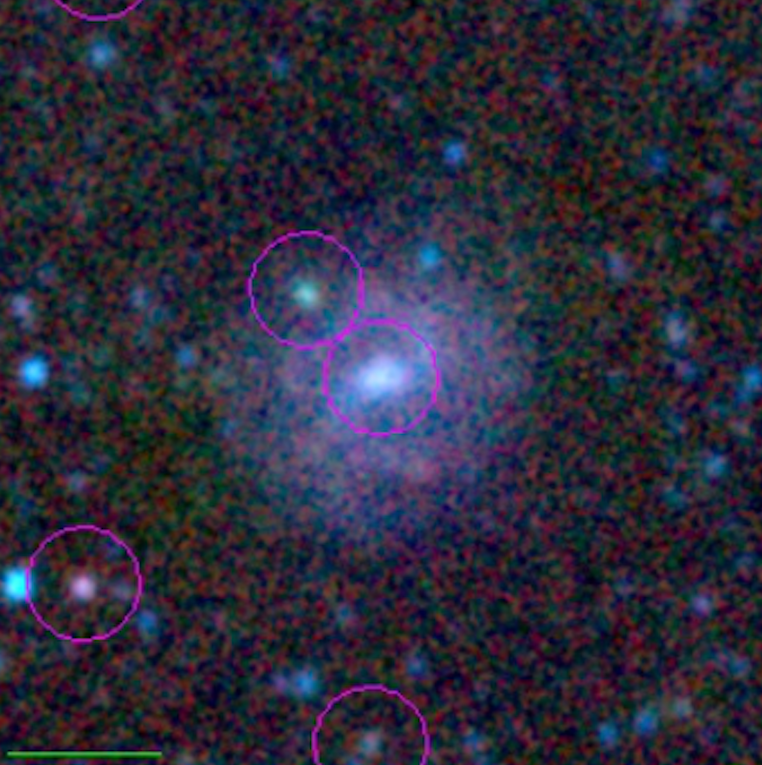}
\end{tabular}
\caption{3.5'x 3.5' SDSS cutouts (left column) of galaxies from the AMIGA-\textit{WISE} sample with the corresponding \textit{WISE} images shown alongside (right column). The green line in each \textit{WISE} image marks a 1' scale, magenta ellipses mark the location of the primary galaxy in each image, as well as any potential galaxy candidates in the field. }
\label{fig:amigas}
\end{figure*}

\subsection{HI pair sample}

\begin{figure*}
\begin{tabular}{|c|c|}
\includegraphics[height = 70mm]{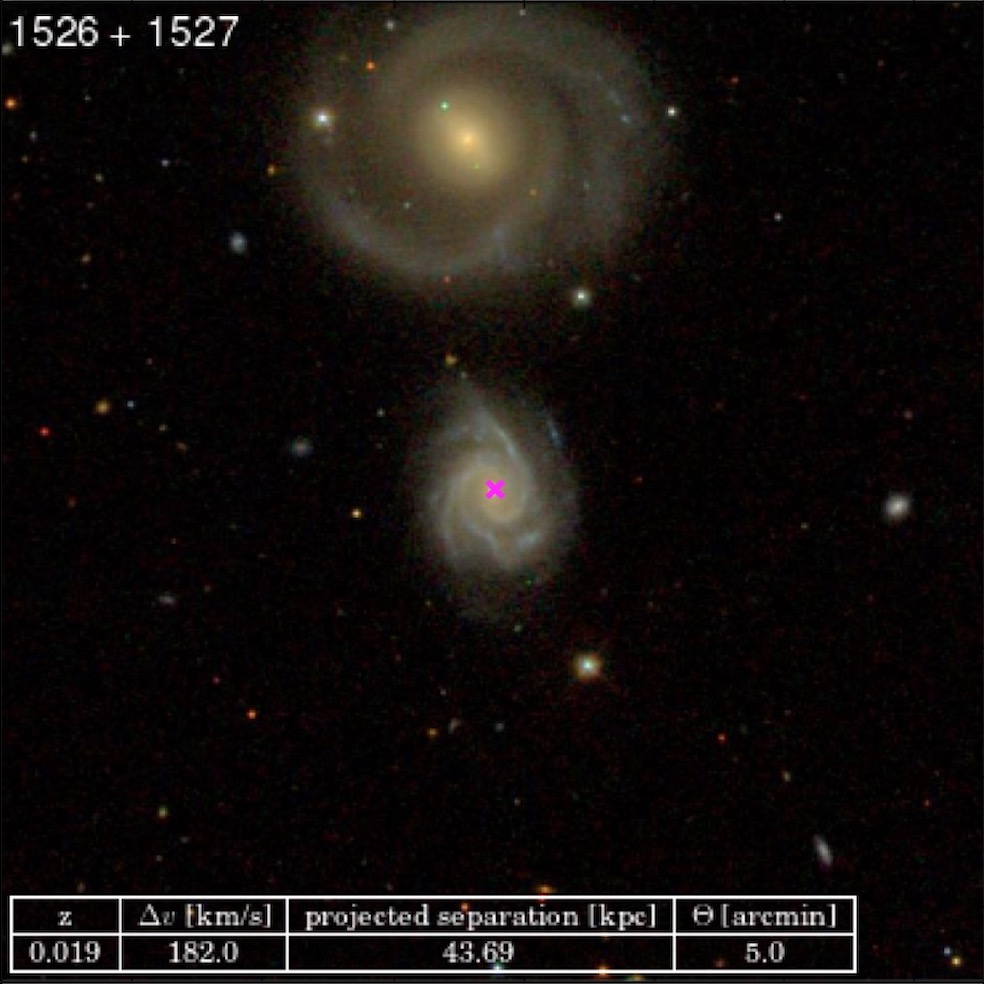}&
 \includegraphics[height = 70mm]{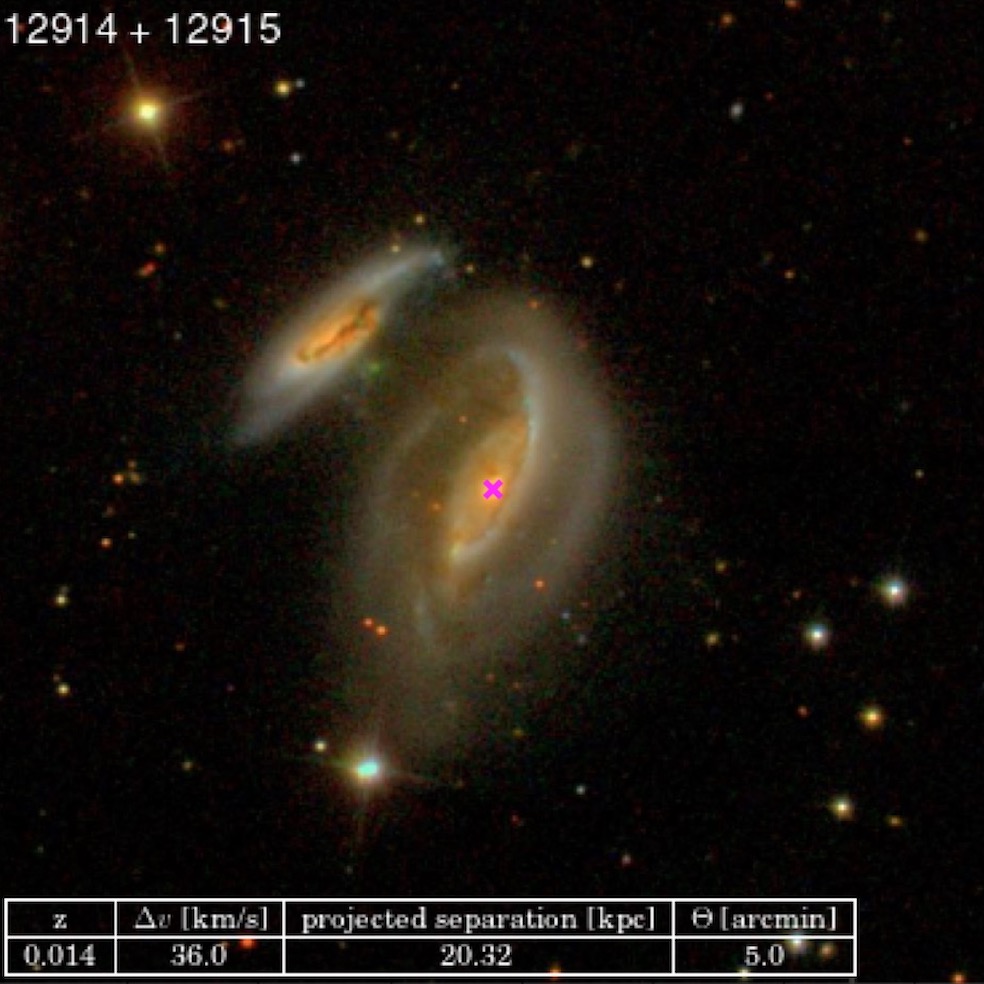} \\
  \includegraphics[height = 70mm]{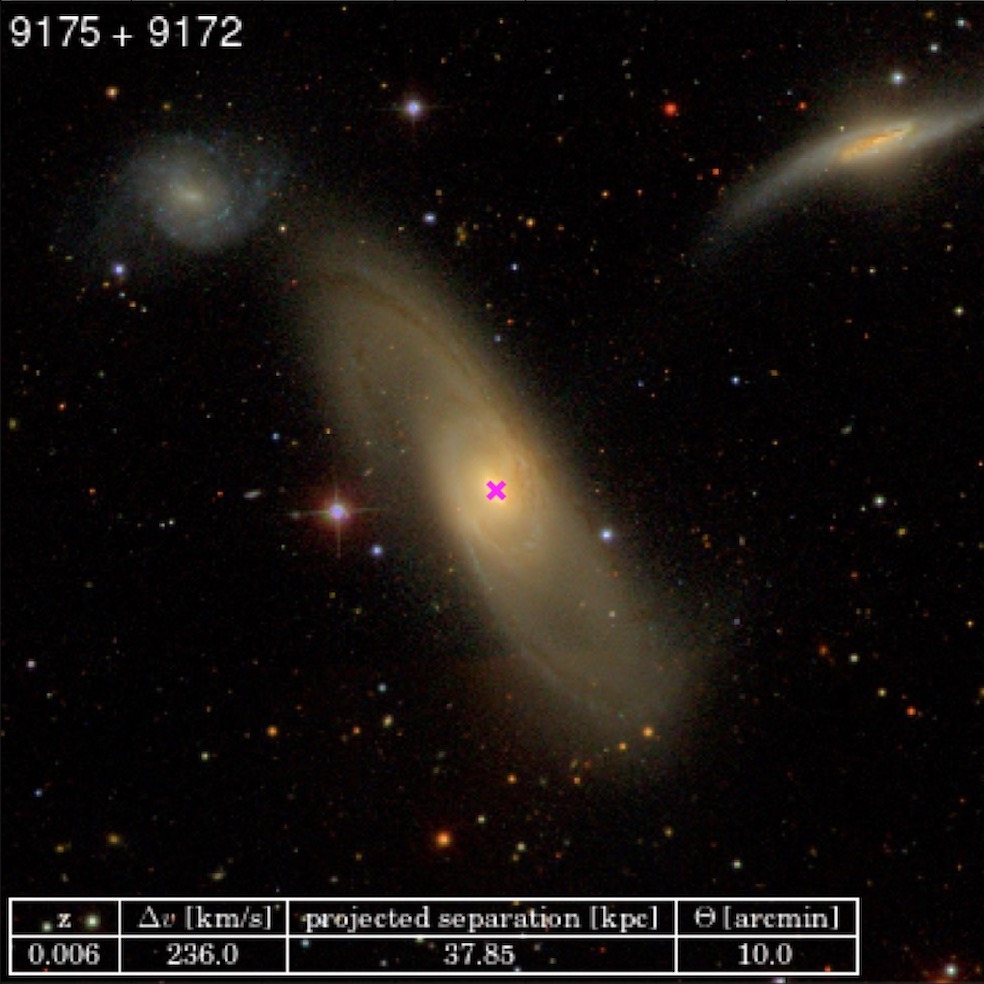}&
  \includegraphics[height = 70mm]{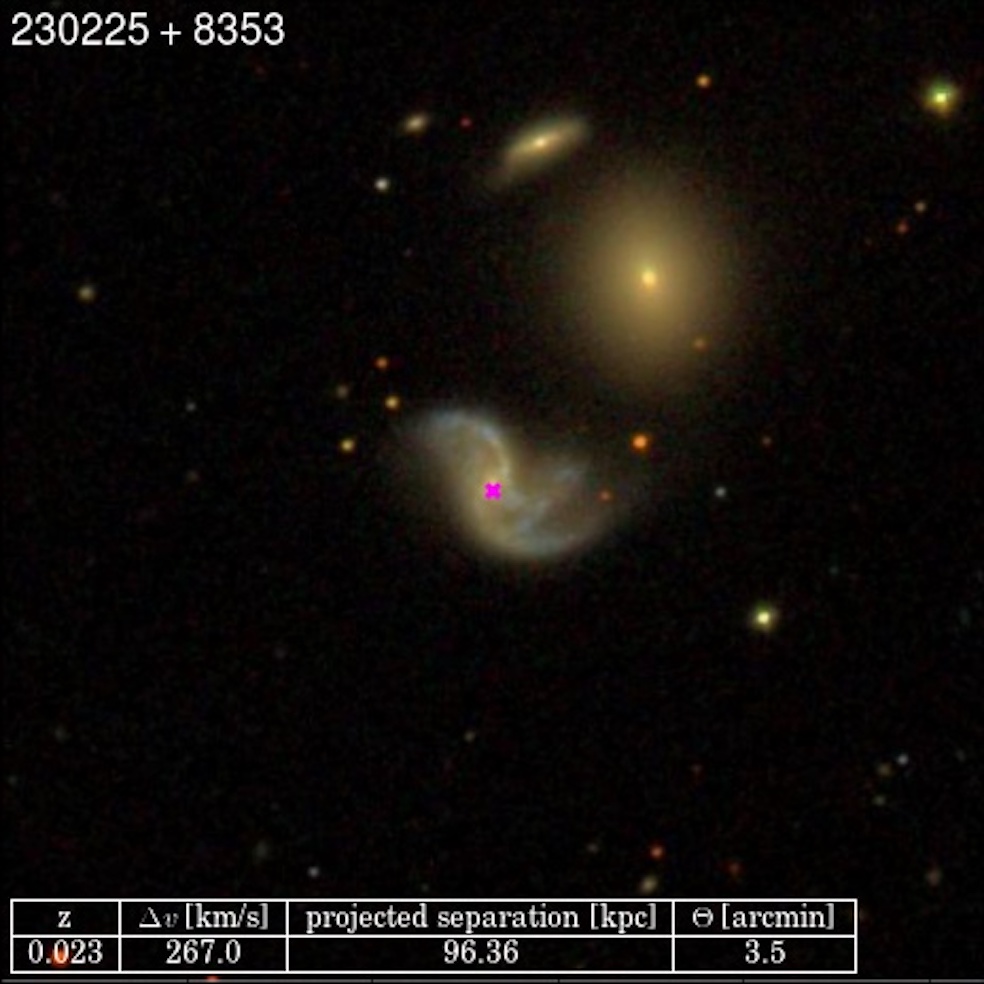}
 \end{tabular}
 \caption{SDSS cutouts of 4 galaxy pairs from the HI pair sample. In each image the redshift, velocity [\kms] and projected [kpc] separation between the pair members, as well as the field of view of the image, are tabulated in the bottom left corner. We note that the galaxy pairs shown in the bottom panel are in fact members of groups.}
\label{fig:pair_sample}
\end{figure*}

For an environmental comparison to our isolated galaxy sample we construct a catalogue of close HI pair galaxies from the Arecibo Legacy Fast ALFA (ALFALFA) survey \citep{Giovanelli2005}, namely the $\alpha 70$ catalogue, for which integrated HI profile masses are also available. Details of the catalogue are discussed in \cite{Haynes2011} and \cite{Haynes2018}. We use projected ($\Delta r$) and velocity ($\Delta v$) separations  to identify a sample of potentially interacting HI galaxies to compare the impact of interactions on gas supply and SF on the SFR/M$\star$ sequence. We do not exclude triples, compact groups, and groups, but rather impose a minimum environment condition that each galaxy in our sample has at least one nearby HI neighbour. We only consider pairs in which both components have a minimum integrated HI profile mass of $10^9$ M$_{\odot}$, marking the steep cut off in HI mass below which ALFALFA detects relatively few galaxies.  
A step-by-step description of our pair finding method follows here: \\

\begin{enumerate}
\item{For each $\alpha 70$ galaxy (M$_{\rm{HI}} > 10^9$ M$_{\odot}$) flagged as being reliably detected and with a spectroscopic counterpart in SDSS DR7 (primary galaxy), we compute the two dimensional distance (in degrees) to its nearest $\alpha 70$ neighbour (secondary galaxy) satisfying the same criteria. As a consequence of this selection criterion our sample is biased towards spiral+spiral pairs, with spiral+elliptical pairs making it into the sample less frequently.}
\item{This distance is then converted to an angular separation using  kpc/'' conversions calculated at the redshift of the primary galaxy.}
\item{The velocity separation is then computed for each pair using spectroscopic redshifts in SDSS. Photometric redshifts are highly uncertain compared to spectrocscopic redshifts, therefore insisting upon spectroscopic redshifts minimizes uncertainty in the calculation of $\Delta v$, and hence the potential for spurious pairs being introduced into our sample.}
\item{We construct our sample of interacting galaxies from those pairs for which the projected separation is less than 100 kpc, and $\Delta v$ is less than 1000 \kms, with a minimum angular separation of 5''. Our $\Delta r$ and $\Delta v$ cuts match the broadest selection used by \cite{Robotham2014} to denote `close pair galaxies'. We note that a $\Delta v$ of 1000 \kms \ is more indicative of galaxies in the same group/large scale structure, however, very few of our pairs have velocity separations exceeding 800 \kms, with the majority of the sample having $\Delta v <$ 400 \kms.}

\end{enumerate}

We find a sample of 282 gas rich pairs in total, in which there are 7 incidents of multiplicity (galaxies paired with more than one galaxy), and therefore 557 unique pair members. As a result of the requirement that both pair candidates have M$_{\rm{HI}} > 10^9$ M$_{\odot}$, the resulting pair sample comprises mainly systems with stellar mass ratios $\sim1$. A quick qualitative visual inspection of the HI pair sample demonstrates that it is similarly spiral galaxy dominated, however with the inclusion of notably more interesting and irregular morphologies as well. We also note a range of local environments including isolated pairs, triples, and compact group candidates. While this paper focusses solely on differences in quantities measured between isolated galaxies and a control sample of galaxies that are in close arrangements (primarily pairs), we will address in detail the role of local environment in an upcoming paper where we quantitatively establish the local environments of our HI pair sample. Optical and MIR images of galaxies from the AMIGA sample are shown side by side for reference in Figure \ref{fig:amigas}, and optical images of some example galaxies from our pair sample are shown in Figure \ref{fig:pair_sample}.

\subsubsection{HI blending in the pair sample}
Of the ALFALFA pairs sample approximately 50$\%$ of the galaxies were noted as being in some form of blend with their neighbour(s) when source extraction was performed by the ALFALFA team. Given the approximately 3.5' beam of the Arecibo telescope in the L-band and the close proximity of the pairs considered, this high rate of blending is not unexpected. For approximately 3/4 of the blends the human extractor from the ALFALFA team reported that the sources were separable to some degree, and thus that the sources parameters should not be heavily biased. The ALFALFA team performed this separation either by careful tailoring of the extraction box, both on the plane of the sky and in velocity space, or in more severe cases by interpolating over heavily affected regions of the source profile. The HI spectra of our pairs were all individually inspected and those with significant signs of blending were removed (49 galaxies in total). It should be noted that the removal of blended galaxies was done on an individual basis, rather than considering a pair as a single object. This is because in pairs with large mass ratios, blending can be severe for the smaller member, but almost negligible for the larger member. In these cases, although the HI properties of the smaller member may be unusable, the larger member is still in a pair and therefore belongs in the sample. We do not exclude these galaxies when only their location on the MS is considered as their stellar masses and SFRs, measured by \textit{WISE} with its 6\arcsec (W1) and 9\arcsec (W3) beams, are still valid. We mark the location of these sources on the MS in Figure \ref{MS} by blue squares, and discuss a few representative cases in Appendix B.

\subsection{Mass cuts}
\begin{figure*}
\centering
\includegraphics[scale=0.35]{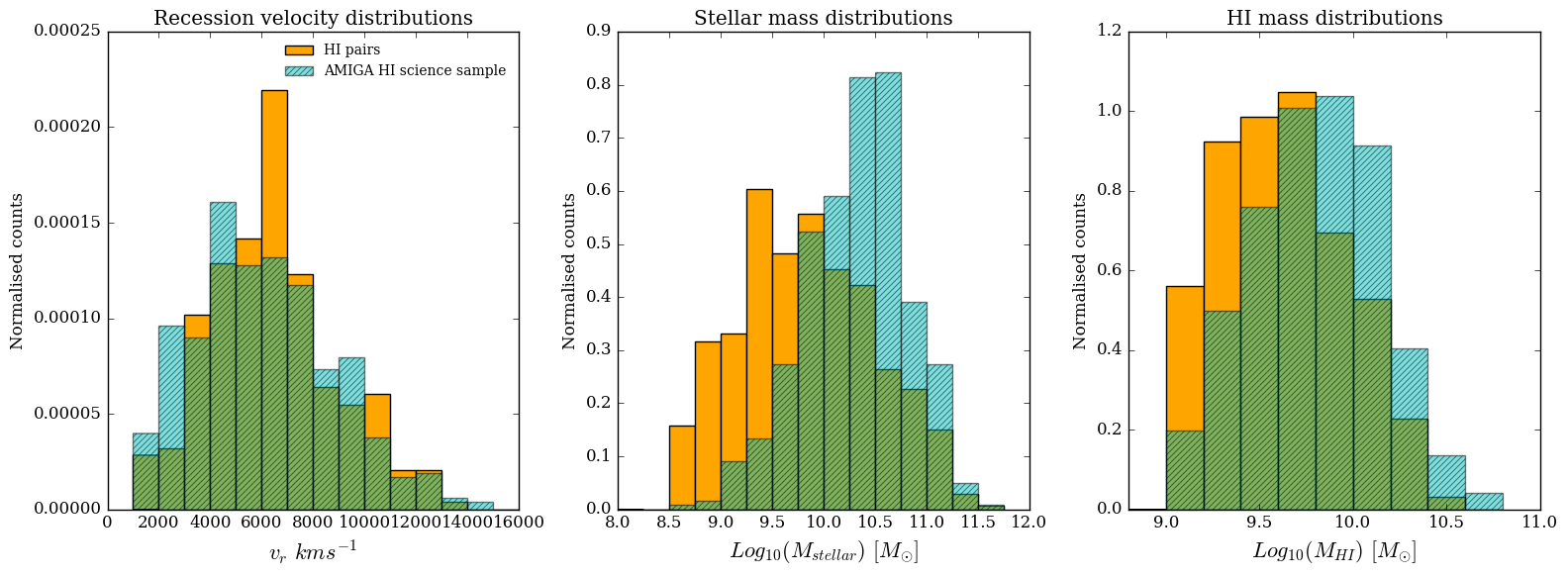}
\caption{From left to right: $v_{r}$, M$\star$, and M$_{\rm{HI}}$ distributions of the AMIGA HI science sample (cyan) and HI pair (orange) samples.}
\label{fig:sample-properties}
\end{figure*}

The recession velocities, M$\star$, and M$_{\rm{HI}}$ distributions of the AMIGA-\textit{WISE} and HI pair samples are shown in Figure \ref{fig:sample-properties}. While both samples probe a similar recession velocity range, we note fewer AMIGA-\textit{WISE} galaxies occupying the lower M$\star$ and M$_{\rm{HI}}$ bins compared to the pair sample. We attribute the missing dwarfs in AMIGA to the particular isolation criterion used in which galaxies below 1500 \kms are excluded so as to avoid searching a prohibitively large area of sky for neighbours (See \cite{Verley2007a} for details on the isolation criteria). Due to their low luminosity, dwarfs are only likely to be detected if they are nearby, excluding nearby galaxies therefore specifically excludes dwarfs from the AMIGA sample. To mitigate the effect of missing dwarfs in our AMIGA sample when we compare them with our pair sample we implement a stellar mass cut of M$_{\star}>10^{8.5}$ \msol \hspace{1mm} across both the AMIGA and HI pair samples. Very few AMIGA galaxies exist below this stellar mass limit, and the HI pair galaxies in this low mass regime correspond to the sources with the largest uncertainties in both their stellar mass and SFR measurements. Exclusion of these low mass sources improves the quality of our study by minimizing uncertainty, and ensuring the two samples are comparable. Our final AMIGA sample comprises 481 galaxies, which we refer to as the AMIGA-\textit{WISE} sample, while 531 galaxy pair members in our HI pair sample survive the stellar mass cut. We further refine and control our analysis by mass-matching our two samples in stellar mass. Our mass-matched samples are created by randomly selecting equal numbers of galaxies in each stellar mass bin. The number of galaxies chosen in each bin is dictated by the sample with the least amount of galaxies available in the bin. The final sample selection criteria are summarized in Tables \ref{tab:amiga_sample} and \ref{tab:pair_sample} for the isolated and pair samples respectively.

\begin{table}
\caption{Size of isolated galaxy sample after each successive cut.}
\label{tab:amiga_sample}
\begin{tabular}{@{}l|l|@{}}
                                                       & Sample size \\ \midrule \midrule
\multicolumn{1}{|l|}{HI Science sample}                & 544         \\ \midrule
\multicolumn{1}{|l|}{\textit{with} WISE measurements}           & 518         \\ \midrule
\multicolumn{1}{|l|}{M$_{\rm{HI}} > 10^9$M$_{\odot}$}   & 482         \\ \midrule
\multicolumn{1}{|l|}{M$_{\star} > 10^{8.5}$M$_{\odot}$} & 481         \\ \bottomrule
\end{tabular}
\end{table}

\begin{table}
\caption{Size of HI pair sample after each successive cut.}
\label{tab:pair_sample}
\begin{tabular}{@{}l|l|@{}}
                                                       & Sample size \\ \midrule \midrule
\multicolumn{1}{|l|}{Close HI pairs }                & 282         \\ \midrule
\multicolumn{1}{|l|}{unique pair members (with M$_{\rm{HI}} > 10^9$M$_{\odot}$)}           & 557         \\ \midrule
\multicolumn{1}{|l|}{M$_{\star} > 10^{8.5}$M$_{\odot}$} & 531         \\ \bottomrule
\end{tabular}
\end{table}

\subsection{\textit{WISE} colours and AGN activity}

\begin{figure*}
    \centering
    
    \includegraphics[width=8cm]{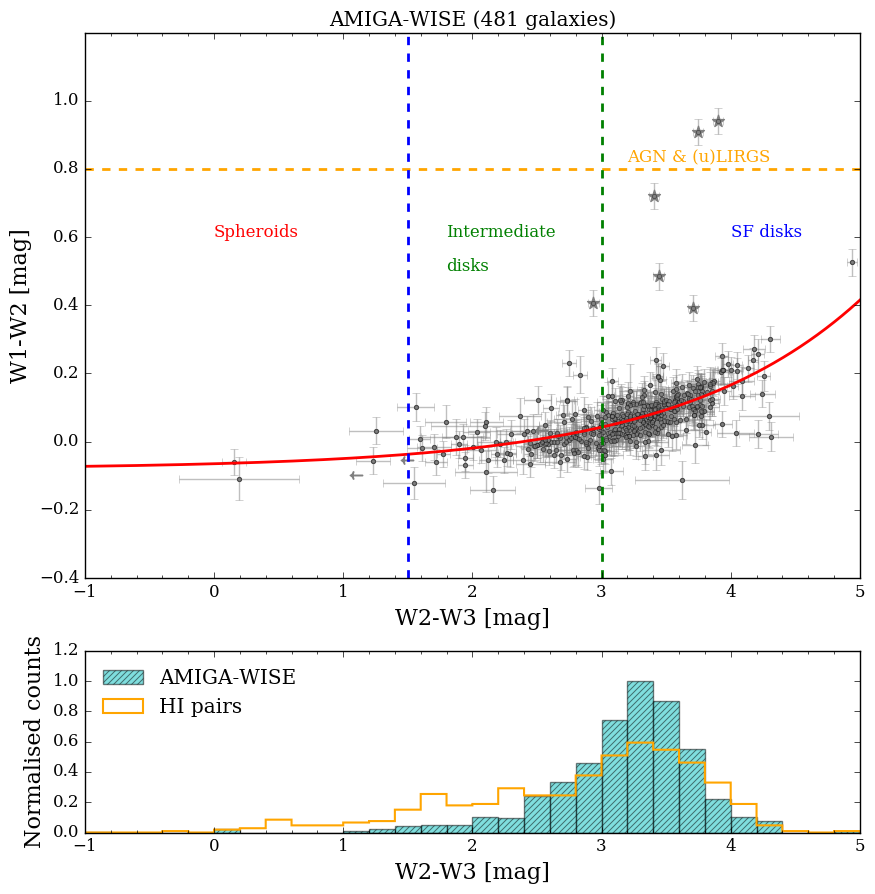}
 
    \qquad
  
    \includegraphics[width=8cm]{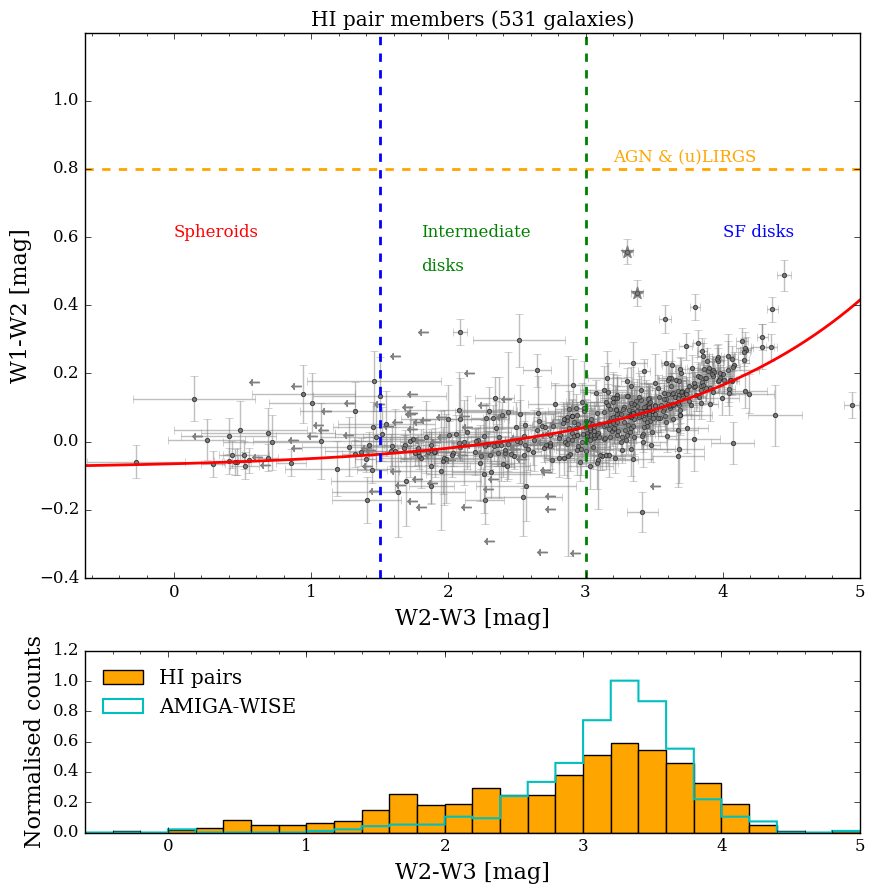}
  
    \caption{\textit{WISE} colour$-$colour diagrams for both the AMIGA-\textit{WISE} (left) and HI pair (right) samples, with the corresponding W2-W3 histogram distributions indicated below each diagram. Upper limits are depicted by gray arrows, and the red curve marks the galaxy MS of Jarrett et al., (2019). MIR AGN candidates are indicated by stars. }
    \label{fig:colour-colour}
\end{figure*}

In Figure \ref{fig:colour-colour} we make use of the \textit{WISE} colour-colour diagram as a diagnostic for estimating galaxy activity in our samples, with which galaxy morphology is largely correlated. (See Section 5.3.3 and Figure 26 in \cite{Jarrett2011} and \cite{Jarrett2019} for a full description of the \textit{WISE} colours and what they indicate.) Gray arrows mark the upper limits, and the galaxy sequence derived by \cite{Jarrett2019} is overlaid in red. The tightness of the AMIGA-\textit{WISE} sample sequence is reflected in the corresponding histogram of W2-W3 color (cyan histogram), which appears to have a single peak in the star-forming disk region of the diagram. In contrast we see more colour diversity in the HI pair sample, not only in the broader distribution and larger scatter of galaxies on the colour-colour diagram, but also in the W2-W3 distribution (orange histogram). The increased scatter we observe in the pair sample relative to AMIGA appears to arise from photometric uncertainties, not to mention systematics that may arise from the fact that the pairs are in complex systems, where deblending from a companion and (if applicable) from other group members is required. The higher frequency of upper limits in the pair sample indicate lower SFRs compared to AMIGA, hinting at the presence of a larger low/no star formation population compared to the AMIGA-\textit{WISE} sample.

In the left panel of Figure \ref{fig:colour-colour} we identify 6 galaxies in the AMIGA-\textit{WISE} sample that are ``warm'' in the MIR  according to their location on the MIR colour-colour diagram (i.e. well above the galaxy sequence). These galaxies are indicated by stars in Figure \ref{fig:colour-colour}. \cite{Sabater2008} classified 5 of these galaxies as AGN using FIR and radio continuum (CIG248, CIG692, CIG993, CIG671, and CIG1004). Of the 2 such candidates identified in the HI pair sample (right panel), NGC 5900 is also a known Seyfert galaxy. The remaining warm MIR galaxies have emission that may be star formation dominated, arise from dust-obscured AGN, or result from a combination of SF and AGN activity. Disentangling SF from AGN activity is non-trivial, however, for the purposes of this work, we conclude that due to the small number of identified AGN candidates, these galaxies are unlikely to affect our results in any substantial way, and thus disregard them as significant contaminators. Furthermore, their locations on the MS in Figure \ref{MS} (black stars) demonstrate that these galaxies are not outliers, and therefore do not skew our results. 

\subsection{\textit{WISE} B/T measurements}
Galaxy morphology is often quantified by decomposing the disk (rotational kinematics, younger populations) and bulge or spheroidal (random motions, older components) and comparing their integrated light properties.  For WISE W1 ($3.4\mu$m) imaging, \cite{Jarrett2019} estimate the bulge and disk light fractions using the axi-symmetric radial distribution,  modeled with a double Sersic profile consisting of the inner bulge and the extended disk components.  For normal galaxies, the resulting near-infrared B/T (bulge-to-total) ratio should provide information on the stellar populations that comprise the bulk mass of the galaxy.  \cite{Jarrett2019}  showed that galaxies with passive or no SF have high ratios, B/T > 0.5, indicating dominant spheroidal populations, and conversely lower-mass star-forming galaxies have low ratios, B/T < 0.3, younger disk-dominated systems, while intermediate-transition galaxies have B/T ratios that range across the spectrum. Due to the large beam of WISE (6''), small and compact galaxies have uncertain decompositions. The  AMIGA-\textit{WISE} and HI pair sample B/T measurements are presented and discussed in Section \ref{BTdiscussion}.

\subsection{\textit{WISE} stellar mass and SFR measurements}

Any inference we make from a galaxy's location on the MS is only as valuable as the SFRs and stellar masses we use are accurate. To this end we make use of mid-infrared M$\star$ and SF indicators in \textit{WISE} (Wide-field Infrared Survey Explorer; \cite{Wright2010} to supply reliable stellar masses and SFRs. The \textit{WISE} W1 $3.4\mu$m and W2  $4.6\mu$m bands trace the stellar mass distribution in galaxies, and the W3 $12\mu$m and W4 $22\mu$m bands are sensitive to the ISM emission from star-forming galaxies, polycyclic aromatic hydrocarbons (PAH) and warm dust respectively \citep{Jarrett2013, Cluver2017}. Using total infrared luminosity ($L_{TIR}$) to calibrate the \textit{WISE} W3 and W4 SFRs, \cite{Cluver2017} found W3 to be particularly good at tracing SFR, measuring a $1\sigma$ scatter in the relation of 0.15 dex over 5 orders of magnitude. Provided deep silicate absorption features and AGN are excluded, the W3 SFR can be considered reliable. The W4 relation has a slightly higher scatter in contrast ($1\sigma$ = 0.18 dex), however both the W3 and W4 relations agree with radio continuum-derived SFRs, as well as the `hybrid' H$\alpha$ and FUV SFR indicators. 
In this paper, stellar masses are computed using a combination of the \textit{WISE} W1 $3.4\mu$m and W2 $4.6\mu$m bands and corresponding mass-to-light ratios from \cite{Cluver2014}, and SFRs are obtained using the W3 $12\mu$m band and \cite{Cluver2017} W3 relation (Equation \ref{w3}).

\begin{equation}
    \rm{Log} SFR(M_{\odot} yr^{-1}) = (0.889 \pm 0.018)~ \rm{Log} L_{12\mu m}(L_{\odot})-(7.76 \pm 0.15)
\label{w3}
\end{equation}

Given the relatively large beam size of \textit{WISE}, blending of galaxies is a general problem, and is addressed accordingly in the \textit{WISE} characterization pipeline. We note that our HI pair sample includes de-blended measurements for 6 galaxy pairs, marked on the MS by red squares in Figure \ref{MS}, which shows them all to be star-forming, and relatively high mass. Adequate de-blend solutions are measured for these galaxies, which are well within the scatter of SF-MS, and therefore not excluded from our analysis. Full details of the \textit{WISE} de-blending process in general, as well as on a case-by-case basis, are discussed in Appendix A. \\
A comparison of the \textit{WISE} stellar masses measured for our isolated sample and optical stellar masses calculated by \cite{FernandezLorenzo2013} show them to be highly consistent, and as such we rule out uncertainty in the stellar masses as a significant source of bias.

\section{An updated AMIGA HI scaling relation}
In the seminal study of HI properties of isolated galaxies, \cite{Haynes1984} computed the HI scaling relation for 324 CIG galaxies. Using optical diameters as a proxy for stellar mass, this relation predicts the HI content of galaxies on the secular evolution track, and has been widely used since to determine the quantity `HI deficiency' defined as: 
\begin{equation}
\rm{DEF = log M_{HI}^{exp} - log M_{HI}^{obs}}
\label{eqn}
\end{equation}
where M$_{\rm{HI}}^{\rm{exp}}$ is the expected HI mass at a given stellar mass, and M$_{\rm{HI}}^{\rm{obs}}$ is the HI mass observed.\\

\begin{figure*}
\includegraphics[scale=0.7]{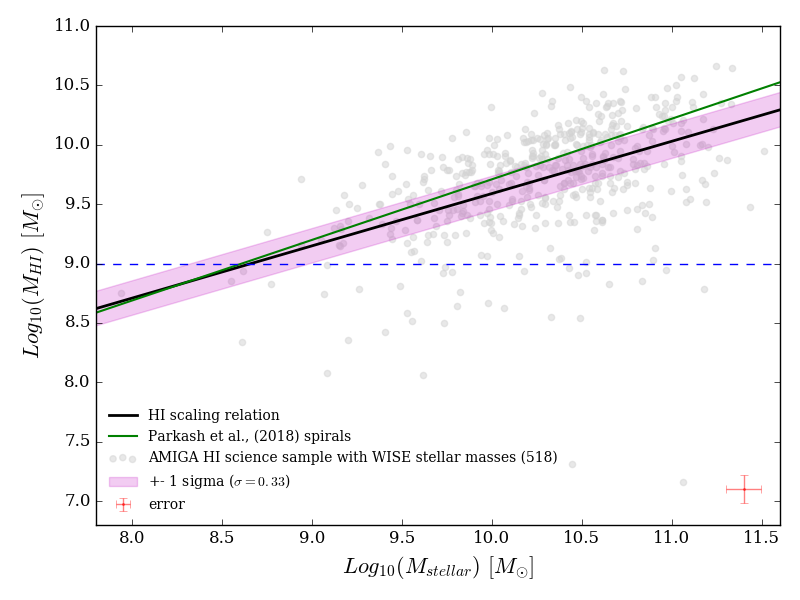}
	\caption{The HI scaling relation (solid black line) for the AMIGA HI science sample galaxies (gray points) with the 1$\sigma$ region shaded in pink. The relation is described by the equation Log$_{10}$(M$_{{\rm{HI}}}$)[\msol] =  0.44 $\times$  Log$_{10}$(M$_{\star}$)[\msol] + 5.19, $\sigma$= 0.33. The dark green line indicates the \protect\cite{Parkash2018} HI scaling relation for spiral galaxies, and the dashed blue line marks the HI mass cut applied to both our isolated and pair samples (M$_{\rm{HI}}>10^9$ \msol). Galaxies below this line are not included in the AMIGA-\textit{WISE} sample.}
	
\label{fig:scaling_relation}
\end{figure*}

\noindent With isolated galaxies providing a baseline for `normal' HI content, computing this quantity for galaxies subject to various different environmental conditions allows one to gauge the impact of environment on HI content. A galaxy's environment might act to deplete it of its HI content, in which case a positive value for HI deficiency is measured, while negative values for HI deficiency indicate an excess of HI is present. A revised HI scaling relation was measured by \cite{Jones2018} using the AMIGA HI science sample. In addition to using a sample superior in both sample size (544 galaxies versus 324) and purity, \cite{Jones2018} also use a more sophisticated regression model to fit the scaling relations. They find that an isolated galaxy's HI content can be predicted by either its optical B-band luminosity or diameter with an accuracy of about 0.25 dex.\\ \\
In this work we update the scaling relation of \cite{Jones2018} using \textit{WISE} W1 $3.4\mu$m stellar mass measurements as an HI predictor. The relationship between stellar and HI mass is more commonly computed in the literature for different galaxy samples. Updating the scaling relation used to compute HI deficiency to rely on stellar mass allows for a direct comparison with various relations in the literature, as well as a direct assessment of how environment impacts HI content.  In Figure \ref{fig:scaling_relation} we illustrate our updated HI scaling relation using \textit{WISE} stellar masses for the AMIGA HI science sample. We fit the HI scaling relation using the maximum likelihood method (for detections) described in \cite{Jones2018}. This method incorporates the measurement uncertainties in both parameters and performs a 3$\sigma$ rejection to remove outliers (which removed the 3 points significantly below the main relation). The resulting relation is described in Equation \ref{fit1} and illustrated in Figure \ref{fig:scaling_relation} as a thick black line. 
\begin{equation}
\rm{Log_{10}(M_{HI})}[M_{\odot}]=0.44 \times \rm{Log_{10}(M_{\star})}[M_{\odot}] + 5.19, \sigma= 0.33
\label{fit1}
\end{equation} 
Pink shading indicates the 1$\sigma$ offset of the relation, and the dark green line marks the fit of \cite{Parkash2018} for a sample of spiral galaxies.  A galaxy's location on the $M_{\star}$/$M_{HI}$ plane above or below the relation suggests either an excess or deficiency in HI content relative to what one would expect of a galaxy in a `nurture free' environment. The scatter in this relation suggests there is some leeway in gas content via consumption and feedback evolution. Our isolated galaxy and pair samples lie above the horizontal dashed blue line as per the HI mass cut (M$_{\rm{HI}}>10^9$\msol).

\section{Results}
\subsection{The SFR/Mstar sequence: Isolated galaxies vs. galaxy pairs}

\begin{figure*}
  \centering
  \subfloat[AMIGA-\textit{WISE}]{\includegraphics[width=0.5\textwidth]{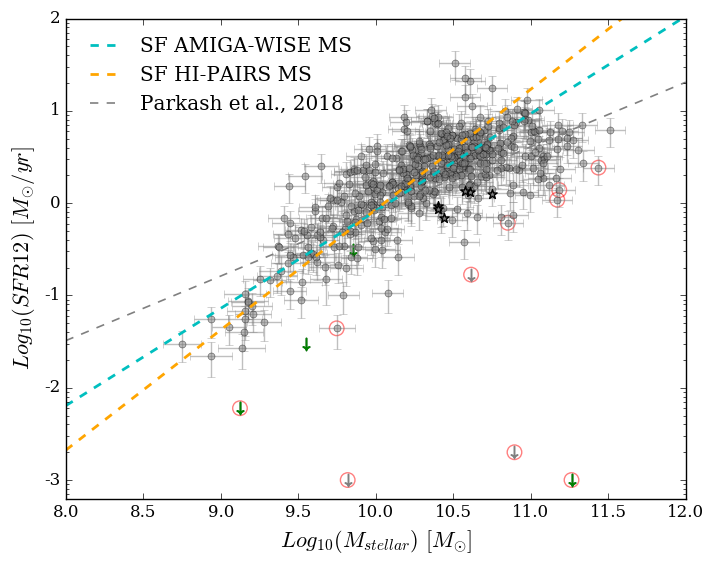}\label{fig:sub1}}
  \hfill
  \subfloat[HI pairs]{\includegraphics[width=0.5\textwidth]{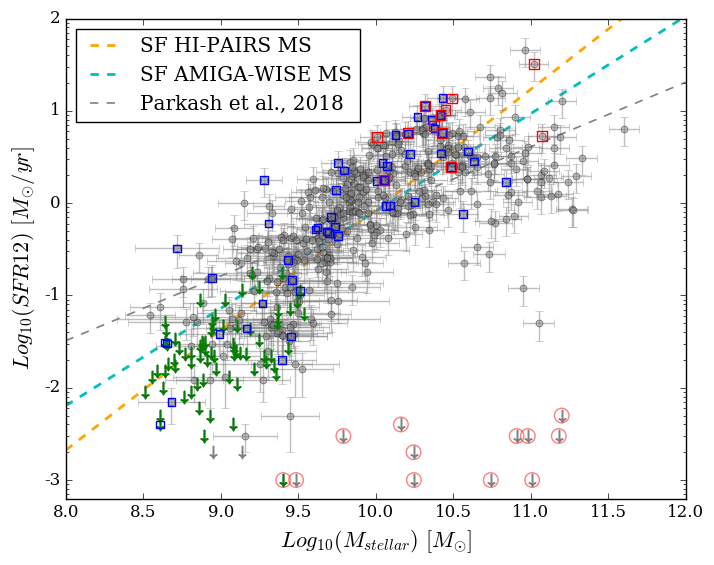}\label{sub2}}
  \caption{The MIR galaxy main sequence. Upper limits on SFR are depicted by downward facing arrows (green indicates no W3 emission has been detected, and gray arrows indicate sources that have been detected in W3, but show no SF). Quiescent galaxies are circled in red, and black stars represent AGN candidates. The AMIGA-WISE, HI pair, and \protect\cite{Parkash2018} MS lines are shown by cyan, orange, and gray dashed lines respectively. Instances of blending in \text{WISE} are marked by red squares, and blue squares identify instances of severe blending in HI that we remove from our HI analysis.}
\label{MS}
\end{figure*}

In Figure \ref{MS} we see the M$\star$/SFR sequence for the full isolated galaxy (left) and pair (right) samples respectively. Upper limits on SFR are marked by gray arrows for sources that have been detected in W3, but show no signs of SF activity. These sources are bright and blue in the MIR, with continuum emission dominating the W3 band. The gray arrows most likely mark galaxies that are old and bulge-dominated in morphology. Green arrows mark the location of sources undetected in W3. No W3 emission in nearby galaxies indicates that the presence of dust re-emission from obscured SF is lacking or minimal, a condition not uncommon for dwarf galaxies.

The AMIGA-\textit{WISE} MS (cyan dashed line) relation is computed in HyperFit for a star-forming sub-sample of the full sample (474 galaxies) excluding upper limits and outliers. The HI pair MS relation (dashed orange line) is computed in the same way for a sample of 454 galaxies. The relations can be found in Equations \ref{amigaMS} and \ref{pairMS} respectively. 
\begin{equation}
\rm{Log}_{10}(SFR_{12})[M_{\odot} yr^{-1}]= 1.05 \times \rm{Log}_{10}(M_{\star})[M_{\odot}] -10.63, \sigma= 0.37
\label{amigaMS}
\end{equation}

\begin{equation}
\rm{Log}_{10}(SFR_{12})[M_{\odot} yr^{-1}] =  1.30 \times \rm{Log}_{10}(M_{\star})[M_{\odot}] -13.10, \sigma= 0.55
\label{pairMS}
\end{equation}
Not only do we observe a flatter MS slope for the AMIGA-\textit{WISE} sample, but also a reduced scatter. The steeper slope of the pair MS is indicating higher SF as a function of stellar mass, and is consistent with the findings of \cite{Ellison2010}.

Focussing specifically on star-forming galaxies, however without any environmental selection, \cite{Speagle2014} report a general MS scatter of $\sim 0.2$ based on 64 MSs from 25 papers in the literature. We propose the significantly larger scatter we measure for our pair sample ($\sigma = 0.55$) is environment driven.

According to the \cite{Bluck2016} criterion for passive galaxies we note a significantly larger number of quenched galaxies in the pair sample compared to our sample of isolated galaxies (45 versus 12). These galaxies, which lie at least 1 dex below the MS relation, are circled in red in Figure \ref{MS}. The larger population of quenched galaxies at high stellar masses in the pair sample perhaps hints at quenching via interaction. These galaxies lie primarily in the spheroid region of the MIR colour diagram with W2-W3 colours $<2$ mag. \textit{WISE} colour images of these sources appear blue due to their emission being W1 $3.4\mu$m dominated. 

\begin{figure*}
\centering
\includegraphics[scale=0.5]{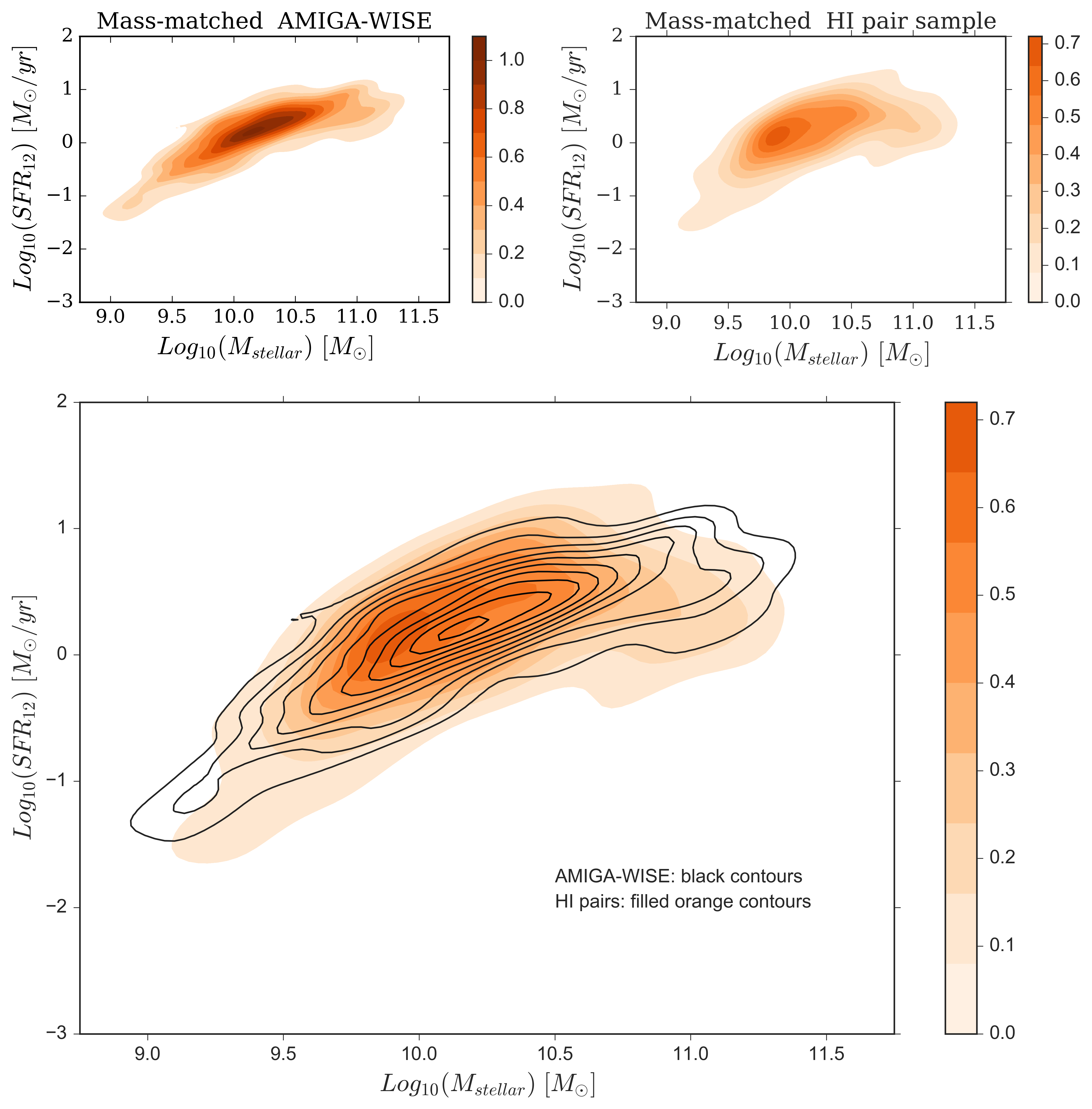}
\caption{Top panel: Filled contour density plots of the AMIGA-\textit{WISE} MS (left) and HI pair MS (right). Bottom panel: Filled contour density plot of the HI pair sample (filled orange contours) with the AMIGA-\textit{WISE} contours overlaid in black.}
\label{fig:kde}
\end{figure*}

In Figure \ref{fig:kde} we plot the density contours of both samples. In the top two panels of the plot we see how the AMIGA-\textit{WISE} sample (left) reaches higher densities compared to the HI pair sample (right), and peaks at a heavier stellar mass. We also note a more pronounced turnover and flattening out of the MS at M$\star \approx 10^{10.5}$ \msol \hspace{0.7mm} in the pair sample (consistent with \cite{Noeske2007} and \cite{Behroozi2013}). In the bottom panel we overlay the AMIGA-\textit{WISE} contours (black contours) on the pair sample filled contour plot for a more direct comparison of the samples. Here we see the AMIGA-\textit{WISE} central density offset to higher stellar masses compared to the pair sample. A visual representation of the decreased scatter in the AMIGA-\textit{WISE} MS (0.37 vs 0.55 for the pairs) is also apparent in the tightness of the AMIGA-\textit{WISE} contours relative to the pair sample. Galaxies in isolation appear to behave in a relatively  predictable manner, tightly confined to the MS. This is often referred to as`secular evolution' \citep{KormendyKennicutt2004}. Introducing companions into the environment appears to produce more stochastic behaviour on the SFR-M$_{\star}$ plane.

\subsection{Gas fraction on the  SFR/Mstar sequence}
\label{gasfraction}
\begin{figure*}
\includegraphics[scale=0.75]{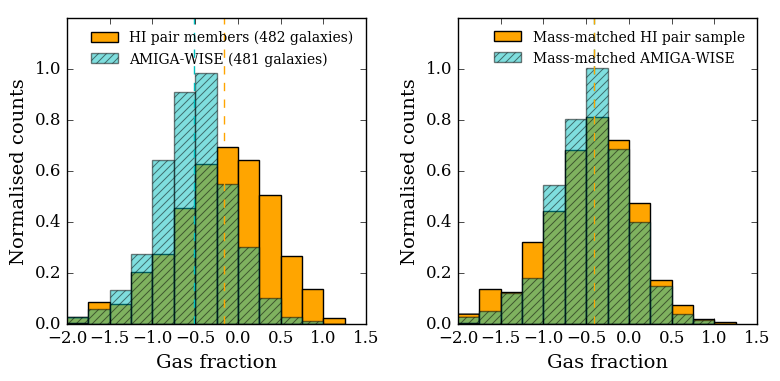} 
	\caption{Gas fraction distributions for the full pair (orange) and AMIGA-\textit{WISE} (cyan) samples are shown on the left, with the corresponding mass matched sample distributions on the right. Vertical dashed orange and cyan lines indicate the respective median gas fraction values for the pair and isolated galaxy samples. These lines merge to occupy the same location on the plot when the samples are matched in stellar mass.}
	\label{fig:gasfraction1}
\end{figure*}

\begin{figure*}
\includegraphics[scale=0.5]{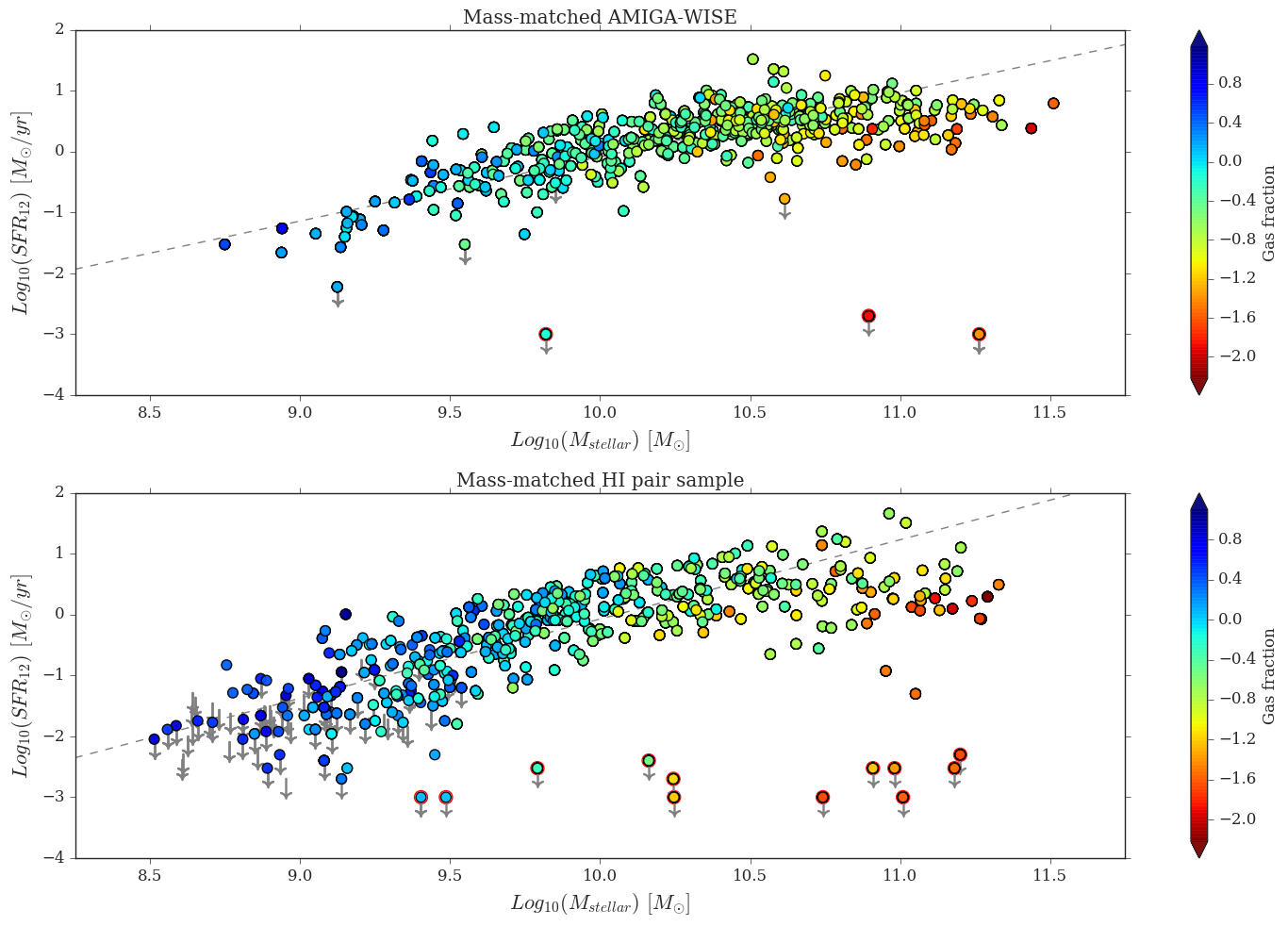} 
	\caption{Enlarged views of the SFR/M$_{\star}$ sequence for both the statistical AMIGA-\textit{WISE} (top panel) and HI pair (bottom panel) mass-matched samples colour coded by gas fraction. Gray arrows indicate upper limits on SFR, of which there are noticeably more in the HI pair sample, concentrated in the lower mass bins. Red circles mark the location of the low star-forming/quenched galaxy candidates. In both samples we see a trend of decreasing gas fraction with increasing stellar mass. We also note in the pair sample an increased availability of galaxies to choose from in the low stellar mass bins in the randomized re-sampling process of mass-matching to the AMIGA sample.}
	\label{fig:gasfraction2}
\end{figure*}

As per Section 2.2.1, 49 instances of severe blending in the HI pair sample are excluded from our HI analysis due to having unreliable HI data. In Figure \ref{fig:gasfraction1} we plot the gas fraction distributions of both the original (left panels) and mass-matched (right) AMIGA-\textit{WISE} (cyan) and HI pair (orange) samples. Here gas fraction is defined as $\rm{Log_{10}(M_{HI}/M_{star})}$. In both panels the cyan histogram refers to the AMIGA-\textit{WISE} sample and the orange histogram corresponds to the pairs, with cyan and orange dashed lines marking the corresponding distribution medians. Focusing first on the left panel, the original AMIGA-\textit{WISE} and HI pair samples, the pair sample appears to be shifted towards higher gas fractions. We measure a median gas fraction value of -0.16 and $\sigma = 0.58$ for the pair sample,  while a significantly lower median and sigma value is measured for the AMIGA-\textit{WISE} sample,  median = $-0.51$ and  $\sigma = 0.43$. We note, however, that the distributions become very similar when the samples are matched in stellar mass (right panel). Here we see very similar medians for both samples, -0.42 and -0.47 for the AMIGA-\textit{WISE} and HI pair samples, respectively. This highlights the importance of controlling for stellar mass as a driver of galaxy properties. The difference between the two mass-matched samples is however evident in the respective widths of the distributions. We can see by eye that the pair sample distribution is slightly broader than the AMIGA-\textit{WISE} distribution, with a larger fraction of the pair sample occupying the higher and lower gas fraction bins. This is reflected by the corresponding measured sigma values: $\rm{\sigma_{pairs} = 0.54}$ vs $\rm{\sigma_{A-W} = 0.44}$. This means for a given stellar mass bin, the pairs show greater variation in gas fraction. Even though there is a scaling relation between stellar mass and HI mass, stellar mass does not appear to constrain the spread of gas fraction. The breadth in gas fraction could therefore be contributing to the increased scatter observed on the pair MS (see Figure \ref{fig:kde}). 

Figure \ref{fig:gasfraction2} shows how gas fraction behaves on the MS. We see similar trends of decreasing gas fraction with increasing stellar mass on the MS for both the AMIGA-\textit{WISE} and HI pair samples. These results suggest that galaxies, regardless of environment, build up their stellar mass content while their HI reservoirs are correspondingly depleted, in agreement with the work of \cite{Saintonge2016}. The sample of \cite{Saintonge2016} lacks any pre-selection against environmental features and the same trend of decreasing gas fraction along the MS is observed. We also note a larger fraction of high gas fraction (HI dominated) sources in the low stellar mass regime of the HI pair sample compared to the AMIGA-\textit{WISE} sample. Our pair sample selection allows for groups of galaxies, and 14 counts of multiplicity confirms the inclusion of triples/small groups (2 or more galaxies paired to the same companion). Visual inspection of the sample suggests the presence of additional potential triples, compact groups, and groups in which additional companions do not satisfy the mass selection criteria of this study. The increased fraction of high gas fraction galaxies in the pair sample may be related to the increased availability of HI in over-dense regions in the cosmic web of HI galaxies. These low mass and HI rich galaxies are missing from the AMIGA-\textit{WISE} sample due to their isolation criterion, which not only eliminates galaxies in pair and group environments, but is biased against nearby dwarf galaxies. In the high stellar mass regime we note that gas fraction alone does not predict a galaxy's location on the MS. High mass galaxies with low gas fractions can be found both on and near the MS, as well as significantly below it in a region we conservatively define as reserved for quenched galaxies (red circles in Figure \ref{fig:gasfraction2}). The majority of the high stellar mass galaxies with low gas fractions found outside the `quenched' region of the MS are located just below the MS, possibly in the process of dropping off the MS to join the quenched population galaxies. The colours of these galaxies, in the act of quenching, are what we call intermediate disks, sometimes referred to as:``green valley''.\\ While gas fraction is widely used in the literature, we note that the relation is not linear, and that a significant residual remains when the $\rm{M_{HI}/M_{\star}}$ ratio is taken. Gas fraction is therefore not the most appropriate quantity to use to determine `normalcy'. We consider HI deficiency is the more robust quantity to use in this regard.

\subsection{HI deficiency on the SFR/Mstar sequence}

\begin{figure*}

\includegraphics[scale=0.65]{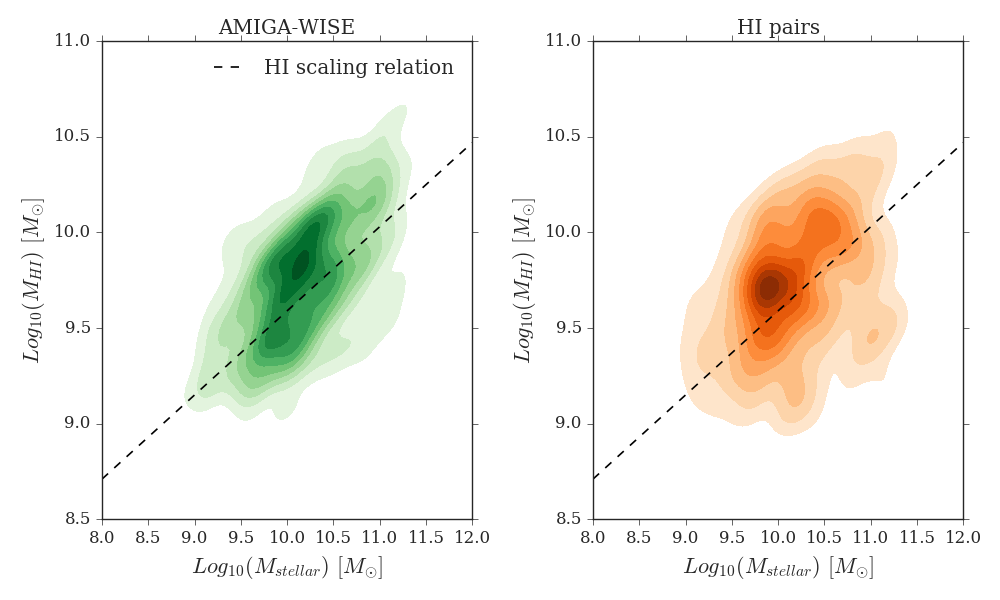}
	\caption{The AMIGA-\textit{WISE} (left) and HI pair (right) samples on the stellar mass-HI mass plane. For reference, the updated AMIGA HI scaling relation (dashed black line, Equation \ref{fit1}) is overlaid. It is important to note that the slope of the HI scaling relation takes into account AMIGA galaxies with M$_{\rm{HI}}<10^9$~\msol\hspace{1mm} that are specifically excluded from the AMIGA-\textit{WISE} sample. The AMIGA-\textit{WISE} sample is thus a gas-rich sub-sample of the AMIGA HI sample, and as such is biased toward galaxies with an excess of HI (located above the relation).}
\label{fig:scaling_relation_both}
\end{figure*}

Figure \ref{fig:scaling_relation_both} shows the stellar mass/HI mass planes for both the AMIGA-\textit{WISE} (left) and HI pair (right) samples with the updated AMIGA HI scaling relation overlaid (black dashed line; Equation \ref{fit1}). The AMIGA-\textit{WISE} galaxies form a tight sequence on the $\rm{M_{\star}/M_{HI}}$ plane as indicated by the tightness of the contours. The paired galaxies form a more irregular sequence in comparison, with fewer tightly spaced contours and a secondary central density potentially indicating two ``flavours'' of paired galaxies on the $\rm{M_{\star}/M_{HI}}$ plane: a low mass sequence and a high mass sequence. Both samples are concentrated above the relation suggesting the presence of excess HI is more frequent than a deficiency of HI in both samples. This result is likely a consequence of the imposed lower limit on stellar and gas mass discussed in Section 2.5, which makes the AMIGA-\textit{WISE} sample a gas-rich subset of the AMIGA sample. 

\begin{figure*}
\includegraphics[scale=0.7]{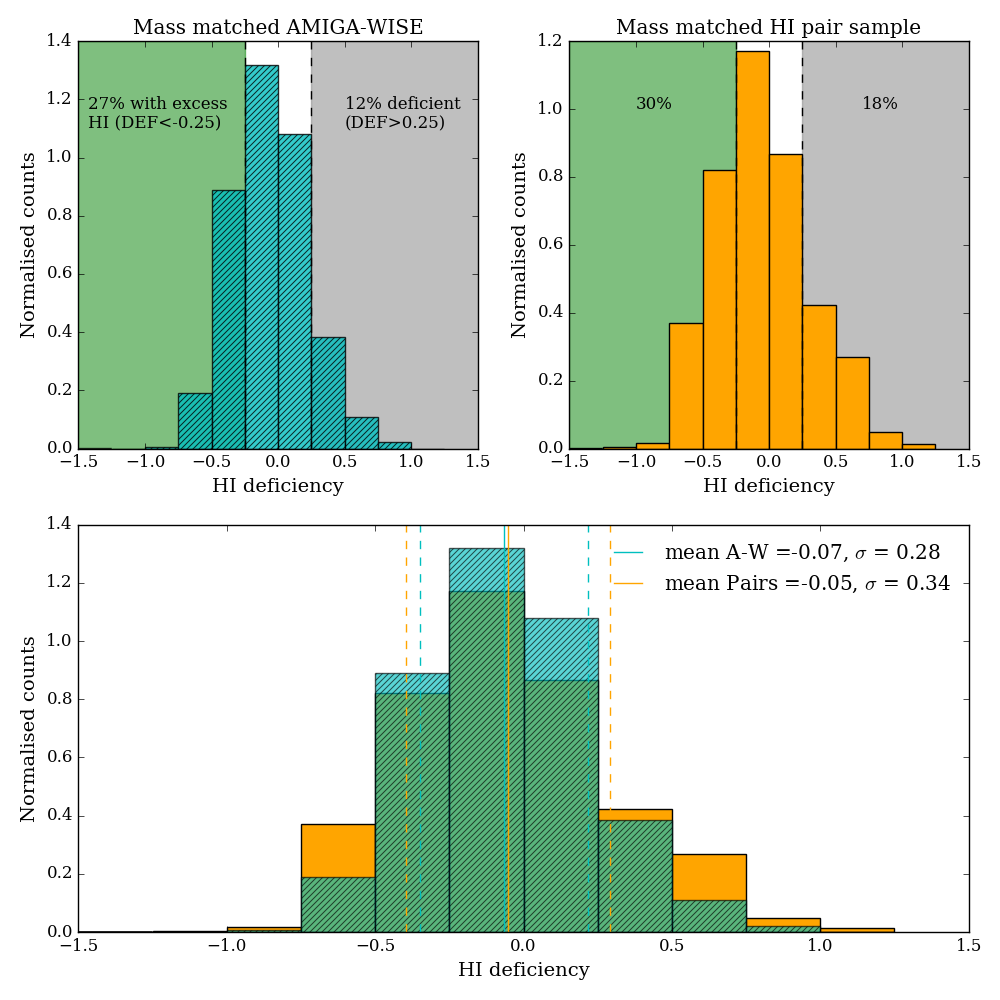}
	\caption{Top: distribution of HI deficiency for the AMIGA-\textit{WISE} (left) and HI pair (right) samples. Gray shading marks the HI deficient regions in each sample, and the green shading shows the region of HI excess. Bottom: AMIGA-\textit{WISE} HI deficiency distribution (cyan) overlaid on the HI pair deficiency distribution. Cyan and orange vertical dashed lines mark the respective distribution widths.}
\label{fig:deficiency_hists}
\end{figure*}

\begin{figure*}
\includegraphics[scale=0.4]{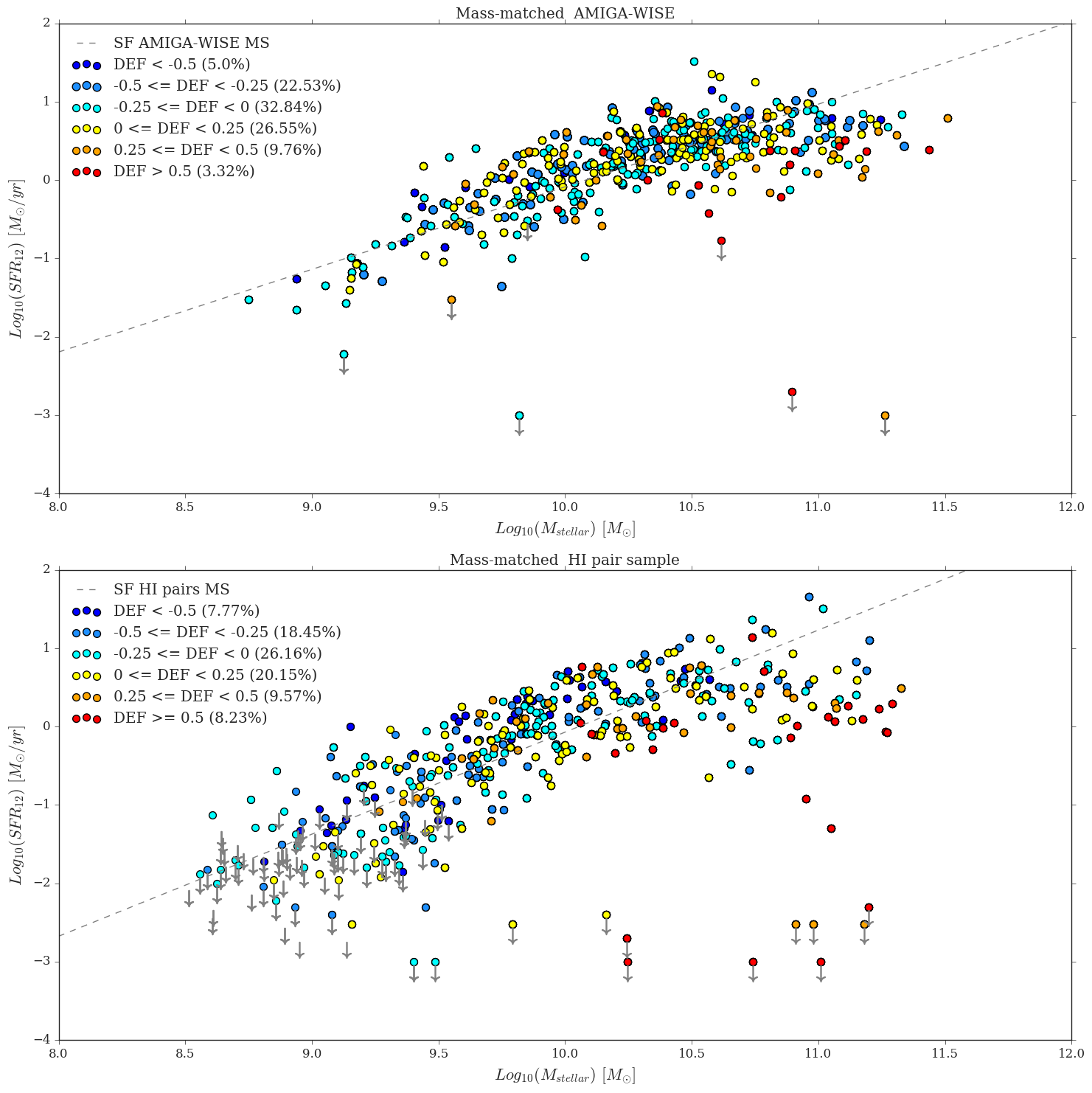}
	\caption{The SFR/M$_{\star}$ sequence for the statistical AMIGA-\textit{WISE} (top) and HI pair (bottom) samples colour coded by HI deficiency.}
\label{fig:HI_deficiency}
\end{figure*}

\begin{figure*}
\includegraphics[scale=0.6]{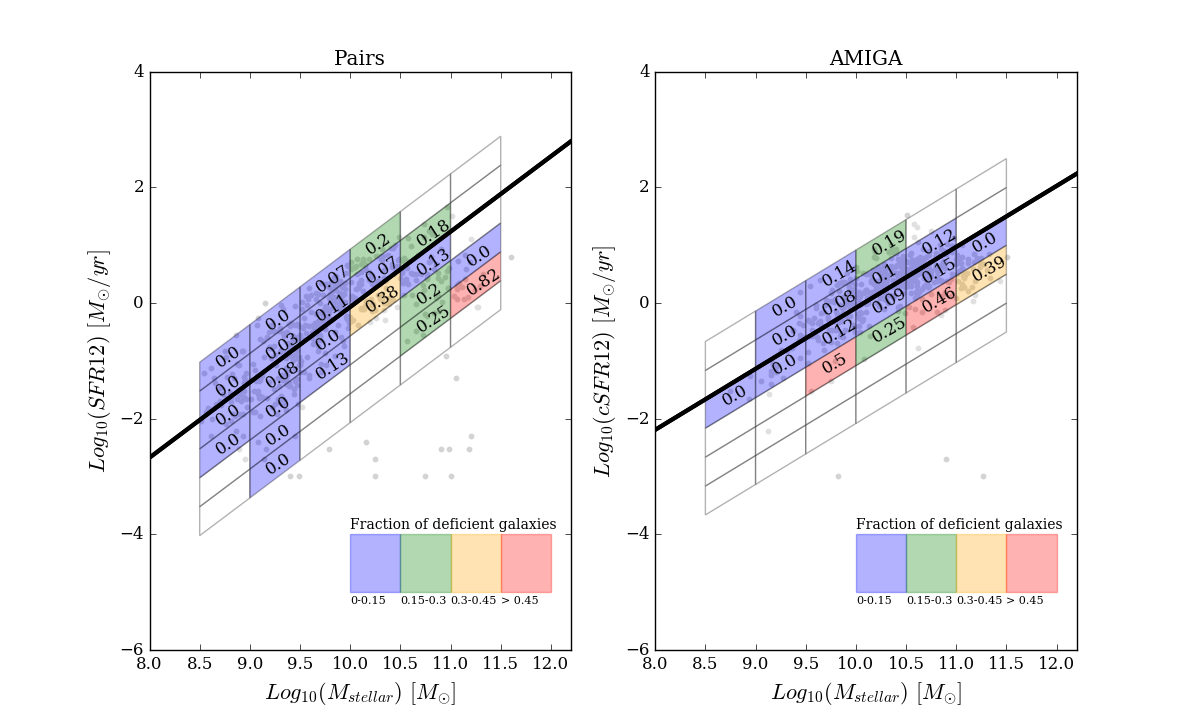}
	\caption{Fraction of HI deficient galaxies (DEF>0.25) on the MS. Here we compare the fractions of deficient galaxies below the MS between the pair (left) and isolated (right) galaxy samples in bins of width 0.5 dex in stellar mass, and height 0.5 dex in SFR. The number in each bin refers to the fraction of galaxies in each bin that measure DEF > 0.25, and the bins are further colour coded as blue, green, orange, or red as per the colour bar key in the bottom right corner of the plot. Bins with less than 2 galaxies are left blank.}
\label{fig:2d_deficiency_bins}
\end{figure*}

In the top panel of Figure \ref{fig:deficiency_hists} we show the distributions of HI deficiency for both the AMIGA-\textit{WISE} (left) and HI pair (right) samples. Here we use the standard definition for deficiency, where positive values indicate a galaxy has less HI than expected given its stellar mass, and lies below the AMIGA HI scaling relation in the M$_{\star}$/M$_{\rm{HI}}$ plane, and negative HI deficiency values indicate an excess of HI. These galaxies lie above the relation. Categorizing the galaxies in each sample as either deficient (DEF>0.25) or in excess of HI (DEF<-0.25), we find a slightly higher fraction of deficient galaxies in the pair sample (18$\%$ compared to 12$\%$ in the sample of isolated galaxies), while both samples have similar fractions of galaxies with an HI excess (27$\%$ and 30$\%$ in the isolated and pair samples respectively). A closer inspection of the deficiency distributions overlaid on each other (bottom panel of Figure \ref{fig:deficiency_hists}), reveals an additional subtle difference between the two distributions. In the inner deficiency bins, from roughly DEF = -0.5 to 0.25, the AMIGA-\textit{WISE} (cyan) distribution lies above the pair (orange) distribution. Conversely, in both the high and low deficiency bins, the pair sample dominates. This is reflected in the width of the distributions. While the samples have very similar mean values (-0.07 vs -0.05 for the AMIGA-\textit{WISE} and pair samples respectively), the width of AMIGA-\textit{WISE} sample (marked by vertical dashed cyan lines) is noticeably narrower than the pair sample (orange dashed lines), with a standard deviation of 0.28 compared to 0.34 for the pair sample. The Anderson-Darling (A-D) test (\cite{Scholz1987}) confirms a statistically significant difference between the samples, and the null hypothesis that the samples are drawn from the same sample can be rejected at the 1$\%$ level. \cite{Jones2018} present a similar comparison between their HI deficiency values for the AMIGA HI science sample and a sample of isolated pairs. They find that the isolated pair sample has more HI deficient galaxies than the AMIGA sample. A similar comparison with galaxies in the Virgo Cluster and the AMIGA sample demonstrates an even larger difference in HI deficiency values, with the Virgo Cluster galaxies exhibiting significantly higher HI deficiency values compared to the AMIGA sample. The findings of \cite{Cortese2011}, \cite{Denes2014}, and \cite{Jones2018}, to name a few, suggest interactions play a role in depleting a galaxy of its HI content. This work suggests environment is responsible for broadening the deficiency distribution of a galaxy sample in both directions, towards higher and lower deficiency values. \\

In Figure \ref{fig:HI_deficiency} we see the AMIGA-\textit{WISE} (top panel) and HI pair sample (bottom panel) SFR/M$_{\star}$ sequences colour coded by HI deficiency. Again we see a a trend of increasing deficiency with stellar mass, but also a broad trend of increasing deficiency with increasing SFR implicating a link between the availability of HI and a galaxy's ability to form stars. This trend is cleaner in the AMIGA-\textit{WISE} sample such that one could almost always predict the rough location of an isolated galaxy on the MS given its HI deficiency value. This is not the case in the pair sample where we observe highly deficient galaxies both on/near the MS, as well as well below it in the quenched galaxy region. Similarly, for lower masses and lower deficiencies, we see galaxies lying both on and below the MS. In both cases one might confuse a star-forming galaxy with a quiescent galaxy if only HI deficiency is considered. This is highly consistent with the picture in which environment drives MS scatter. Focusing specifically on quiescent galaxies, the low mass, high gas fraction population of galaxies discussed in Section \ref{gasfraction} tend to be inefficient at forming at stars despite having an excess of HI (small HI deficiencies). The second population of high mass low gas fraction galaxies have large HI deficiencies. Looking at the fraction of deficient galaxies per stellar mass/SFR bin on the MS in Figure \ref{fig:2d_deficiency_bins}, the predictable nature of the isolated galaxy sample is demonstrated as bands of nearly uniform deficiency fractions. Each bin has a width 0.5 dex in stellar mass and height 0.5 dex in SFR (measured from the respective MS lines), and is colour coded according to the fraction of galaxies with DEF > 0.25 (blue: fraction = 0-0.15, green: fraction = 0.15-0.3, orange: fraction = 0.3-0.45, and red: fraction > 0.45). The lowest band in the isolated sample is the exception with multiple bins of fractional deficiency making up the band, however the fraction is consistently above 0.25 indicating that galaxies more than 0.5 dex below the MS are more likely to be HI deficient. The pair sample, in contrast, shows more variance in each band. Bins with larger fractions of deficient galaxies are spread more liberally above and below the MS, although still more frequently below.   

\section{Discussion}
\label{BTdiscussion}

In section \ref{gasfraction} we note the presence of a quenched population of galaxies in both the AMIGA-\textit{WISE} and HI pair samples. Here we adopt a more conservative definition of quiescence to only include those galaxies that have fallen well below the MS, not only 1 dex below it as per \cite{Bluck2016}. We circle these galaxies in red in Figure \ref{fig:gasfraction2}, 3 in the AMIGA-\textit{WISE} sample and 12 in the HI pair sample. With environment selected to be the controlled variable between the two samples, we propose that the increased number of quenched galaxies found in the pair sample is environment-driven. Visual inspection of the quenched pair members reveals a potentially broader environment at play that includes triples and small groups. 

In both samples we note two flavours of quenched galaxies, those with a low stellar mass and high gas fraction (LM-HG), and those with a high stellar mass and low gas fraction (HM-LG). The LM-HG galaxies also have low HI deficiencies (i.e., gas rich), and the HM-LG galaxies have high HI deficiencies (i.e., gas poor). A likely scenario for the quenching of the latter population of galaxies is simply consumption of fuel. As neutral hydrogen is converted to stars via molecular gas, the HI content of the galaxy decreases while the stellar mass content increases, thus producing high stellar mass galaxies with depleted gas fractions. The increased frequency of quenched galaxies as stellar mass increases we observe is consistent with the findings of Cluver et al., 2020, who make the same observation for  samples of field and grouped galaxies from the Galaxy and Mass Assembly survey (GAMA) with \textit{WISE} photometry, and attribute this result to mass quenching. 

\begin{figure*}
\includegraphics[scale=0.4]{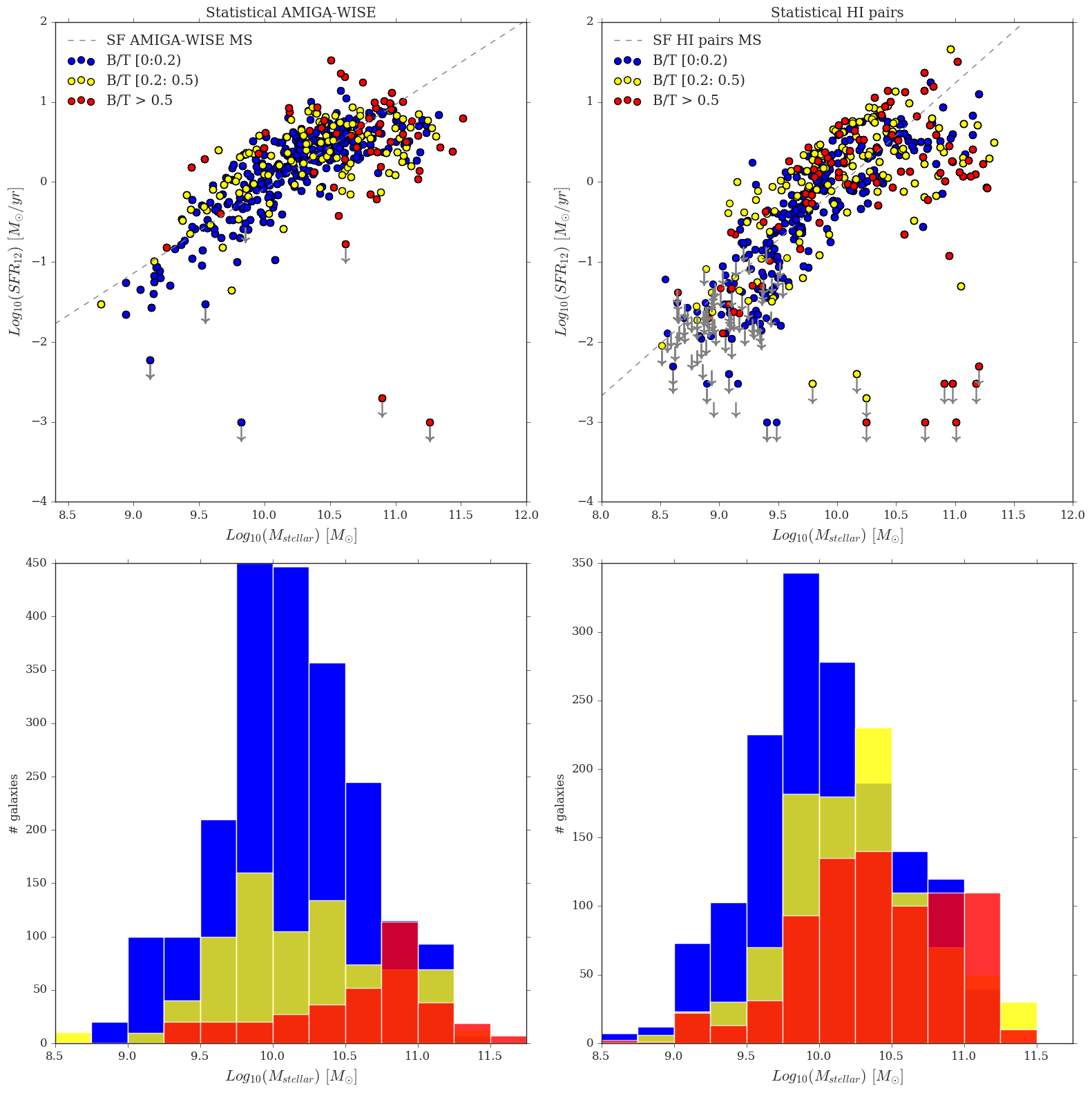}
	\caption{ Top panel: the AMIGA-\textit{WISE} (left) and HI pair (right) sample MS diagrams colour coded by B/T in bins of [0:0.2), [0.2:0.5), and B/T>0.5 (indicated by blue, yellow and red respectively). Bottom panel: stellar mass distributions in the above mentioned B/T bins for the AMIGA-\textit{WISE} (left) and HI pair sample (right).}
\label{fig:BT_amiga}
\end{figure*}

Considering morphology, in Figure \ref{fig:BT_amiga} we look at the bulge to total (B/T) ratios of our sample galaxies. In the top panel of Figure \ref{fig:BT_amiga} we plot the MS for both samples in bins of B/T, blue (B/T<0.2), yellow (0.2<B/T<0.5), and red (B/T >0.5). In the bottom panel we look at the stellar mass distributions of both samples in these same B/T bins. We note that the AMIGA-\textit{WISE} sample (left) is primarily comprised of low B/T disky galaxies populating the full stellar mass range, with only a small portion of the sample in the high mass range measuring large B/T values. These results are consistent with the work of \cite{Durbala2008} and \cite{FernandezLorenzo2013}, who demonstrated a prevalence of pseudo-bulges with low B/T values in sub-samples of AMIGA galaxies. These studies made use of optical photometry, namely the CAS parameters and GALFIT to quantify bulge properties. This suggests optical and MIR B/T measurements are in good agreement.
In contrast to the isolated galaxies, a larger fraction of the pair sample has large B/T (bulgy galaxies) values, spanning a broader range of stellar mass values. The high mass quenched galaxies in particular have large bulges, supporting a possible morphological quenching scenario. In this scenario the formation of a bulge stabilizes the disk against the gravitational collapse necessary for SF to occur \citep{Bluck2014}. \cite{Cook2019}, however, find little evidence to suggest that the presence of a large bulge can alter the gas disk, and cautions interpreting the link between large bulges and quenched galaxies as causal. They suggest quenching processes are more likely occur at the source of in-flowing gas.

The lower mass, low SFR galaxies are disk dominated, gas rich, with low HI deficiencies (excess HI), suggesting an alternative quenching mechanism in the low stellar mass regime. While the fuel for SF is abundant, the column density of the HI gas may be insufficient to ignite SF. The lack of SF in these galaxies might also be as a result of gravitational shock heating of their gas \citep{Liu2017}. Alternatively, these galaxies have simply yet to reach the MS. While HI content is clearly important in moving galaxies along the MS, it does not necessarily account for the distance travelled by galaxies above and below it. Future work will therefore include an investigation into the SF efficiency of our galaxies. With gas rich galaxies found below the MS, and HI deficient galaxies found above it, the efficiency of converting gas into stars is an important quantity to consider next.
 
In the  case of paired galaxies, the influence of other galaxies, in the broader/large scale environment, such as harassment, strangulation, ram pressure stripping, and merging may be disrupting the flow of gas and suppressing SF in both the low and high mass quenched galaxies. The pair sample will be more closely examined in a future paper, which will include a visual inspection for signs of interaction, an analysis of the impact of multiplicity, and a quantitative assessment of the range of local environments using the AMIGA tidal influence and local number density parameters (Q and $\eta$). By establishing the broader pair environment more accurately and quantitatively we will obtain a clearer picture of the role of environment not only contributing to the suppression/enhancement of SF, but in driving the observed scatter in the MS, potentially unlocking important pathways of galaxy evolution. 

\section{Summary}
This paper probes the role of both environment and HI content (gas fraction and HI deficiency) in determining a galaxy's location on the MIR  SFR-M$\star$ sequence. We summarize our main results here:
\begin{enumerate}
    \item{We update the AMIGA HI scaling relation for isolated galaxies of \cite{Jones2018} using \textit{WISE} stellar masses on the x-axis as opposed to the previously used optical B-band diameters and luminosities. This relation, Equation \ref{fit1}, can be used as a baseline predictor of HI content in the absence of environmental influence, and we encourage its use in establishing the impact of environment on HI content by comparing it with galaxy samples in different environments.}
    \item{We compare the SFR-M$\star$ sequence of the AMIGA-\textit{WISE} HI sample with a sample of gas rich HI pairs and compute the MS fit for both samples using HyperFit (Equations \ref{amigaMS} and \ref{pairMS}). We observe a flatter slope for the AMIGA-\textit{WISE} sample, as well as a reduced scatter compared to the HI pair sample ($\sigma = 0.37$ versus $\sigma = 0.55$). We attribute the tightness of the AMIGA-\textit{WISE} MS to the isolated neighbourhood in which these galaxies are located, whereas the more stochastic behaviour of the HI pairs is likely the result of the more densely populated environments in which they live.}
    \item{We compare the mass matched AMIGA-\textit{WISE} gas fraction distribution to that of the HI pair sample and find that while the distribution medians are essentially identical, the significant difference between the two distributions lies in the width. The pair sample gas fraction distribution is notably broader by eye, and quantitatively measures a larger scatter ($\sigma = 0.54$ versus $\sigma = 0.44$ for the AMIGA-\textit{WISE} sample). We also track the behaviour of gas fraction on both the AMIGA-\textit{WISE} and HI pair sample MS. In both samples we observe a trend of decreasing gas fraction with increasing stellar mass. This suggests galaxies build up their stellar mass content via the consumption of HI fuel regardless of the environment in which they reside. The relationship between gas fraction and SFR, however, is non-linear, and does not reliably predict the location of a galaxy on the MS plane. High stellar mass galaxies with similarly low gas fractions are found both on the SF sequence and well below it.}
    \item{The width of the AMIGA-\textit{WISE} HI deficiency distribution is noticeably narrower than that of the pair sample, with a smaller standard deviation of $\sigma = 0.28$ compared to $\sigma = 0.34$ in the pairs. We also note that the pair sample dominates in both the high and low HI deficiency bins, suggesting the potential role of environment simultaneously increasing and depleting HI content. The Anderson-Darling test confirms a statistically significant difference between the samples.    
  } 
  
 \item{MIR B/T values measured in \textit{WISE} show the AMIGA-\textit{WISE} sample to be dominated by low B/T disky galaxies,  in agreement with optical B/T measurements conducted by \cite{Durbala2008} and \cite{FernandezLorenzo2013}. In comparison, a higher frequency of large B/T values is observed in the pair sample, with some of the largest B/T values belonging to the high mass quiescent population galaxies.
 }
\end{enumerate}

The AMIGA sample has previously been shown to have the lowest measured dispersion in various quantities (HI profile asymmetry \citep{Espada2011}, FIR emission \citep{Lisenfeld2007}, and g-r colour \citep{FernandezLorenzo2013}) when compared to galaxies selected without isolation criteria considerations. Similarly, this work shows isolated galaxies to have a lower dispersion in both gas fraction and HI deficiency in comparison to a sample of galaxies in denser environments, as well as a reduced scatter on the SFR-M$\star$ sequence. 
 
\section*{Data availability}
The data underlying in this paper will be made available by request from J. Bok (email: jamie@ast.uct.ac.za). 

\section*{Acknowledgements}
We thank the anonymous referee for their very useful comments which resulted in a greatly improved paper. 
This work is based on the research supported in part by the National Research Foundation of South Africa (Grant Numbers UID: 101099 and 111745). JB additionally acknowledges
support from the DST-NRF Professional Development Programme (PDP), and the University of Cape Town. MC is a recipient of an Australian Research Council Future Fellowship (project number FT170100273) funded by the Australian Government. THJ acknowledges funding from the National Research Foundation under the Research Career Advancement and South African
Research Chair Initiative programs, respectively.
We acknowledge the work of the entire ALFALFA team for observing, flagging and performing signal extraction. MGJ is supported by a Juan de la Cierva formaci\'{o}n fellowship (FJCI-2016-29685) from the Spanish Ministerio de Ciencia, Innovacion y Universidades (MCIU). MGJ also acknowledges support from the grants AYA2015-65973-C3-1- R (MINECO/FEDER, UE) and RTI2018-096228-B-C31 (MCIU).  This work has been supported by the State Agency for Research of the Spanish MCIU through the ‘Centro de Excelencia Severo Ochoa’ award to the Instituto de Astrof\'{i}sica de Andaluc\'{ı}a (SEV-2017-0709).
This publication makes use of data products from the Wide-field
Infrared Survey Explorer, which is a joint project of the University of California, Los Angeles, and the Jet Propulsion
Laboratory/California Institute of Technology, funded by the National Aeronautics and Space Administration.
This work also utilizes data from Arecibo Legacy Fast ALFA (ALFALFA) survey data set obtained with the Arecibo L-band Feed Array (ALFA) on the Arecibo 305m telescope. Arecibo Observatory is part of the National Astronomy and Ionosphere Center, which is operated by Cornell University under Cooperative Agreement with the U.S. National Science Foundation. Funding for the SDSS and SDSS-II has been provided by the Alfred P. Sloan Foundation, the Participating Institutions, the National Science Foundation, the U.S. Department of Energy, the National Aeronautics and Space Administration, the Japanese Monbukagakusho, the Max Planck Society, and the Higher Education Funding Council for England. The SDSS Web Site is http://www.sdss.org/. In addition, we make use of data from the Sloan Digital Sky Survey (SDSS DR7). The SDSS is managed by the Astrophysical Research Consortium for the Participating Institutions. The Participating Institutions are the American Museum of Natural History, Astrophysical Institute Potsdam, University of Basel, University of Cambridge, Case Western Reserve University, University of Chicago, Drexel University, Fermilab, the Institute for Advanced Study, the Japan Participation Group, Johns Hopkins University, the Joint Institute for Nuclear Astrophysics, the Kavli Institute for Particle Astrophysics and Cosmology, the Korean Scientist Group, the Chinese Academy of Sciences (LAMOST), Los Alamos National Laboratory, the Max-Planck-Institute for Astronomy (MPIA), the Max-Planck-Institute for Astrophysics (MPA), New Mexico State University, Ohio State University, University of Pittsburgh, University of Portsmouth, Princeton University, the United States Naval Observatory, and the University of Washington.

\bibliographystyle{mnras}

\bibliography{correct_paper1}

\begin{thebibliography}{}
\makeatletter
\relax
\def\mn@urlcharsother{\let\do\@makeother \do\$\do\&\do\#\do\^\do\_\do\%\do\~}
\def\mn@doi{\begingroup\mn@urlcharsother \@ifnextchar [ {\mn@doi@}
  {\mn@doi@[]}}
\def\mn@doi@[#1]#2{\def\@tempa{#1}\ifx\@tempa\@empty \href
  {http://dx.doi.org/#2} {doi:#2}\else \href {http://dx.doi.org/#2} {#1}\fi
  \endgroup}
\def\mn@eprint#1#2{\mn@eprint@#1:#2::\@nil}
\def\mn@eprint@arXiv#1{\href {http://arxiv.org/abs/#1} {{\tt arXiv:#1}}}
\def\mn@eprint@dblp#1{\href {http://dblp.uni-trier.de/rec/bibtex/#1.xml}
  {dblp:#1}}
\def\mn@eprint@#1:#2:#3:#4\@nil{\def\@tempa {#1}\def\@tempb {#2}\def\@tempc
  {#3}\ifx \@tempc \@empty \let \@tempc \@tempb \let \@tempb \@tempa \fi \ifx
  \@tempb \@empty \def\@tempb {arXiv}\fi \@ifundefined
  {mn@eprint@\@tempb}{\@tempb:\@tempc}{\expandafter \expandafter \csname
  mn@eprint@\@tempb\endcsname \expandafter{\@tempc}}}

\bibitem[\protect\citeauthoryear{{Argudo-Fern{\'a}ndez}
  et~al.,}{{Argudo-Fern{\'a}ndez} et~al.}{2013}]{Argudo-Fernandez2013}
{Argudo-Fern{\'a}ndez} M.,  et~al., 2013, \mn@doi [\aap]
  {10.1051/0004-6361/201321326}, \href
  {https://ui.adsabs.harvard.edu/abs/2013A&A...560A...9A} {560, A9}

\bibitem[\protect\citeauthoryear{{Behroozi}, {Wechsler}  \&
  {Conroy}}{{Behroozi} et~al.}{2013}]{Behroozi2013}
{Behroozi} P.~S.,  {Wechsler} R.~H.,   {Conroy} C.,  2013, \mn@doi [\apj]
  {10.1088/0004-637X/770/1/57}, \href
  {https://ui.adsabs.harvard.edu/abs/2013ApJ...770...57B} {770, 57}

\bibitem[\protect\citeauthoryear{{Bitsakis} et~al.,}{{Bitsakis}
  et~al.}{2019}]{Bitsakis2019}
{Bitsakis} T.,  et~al., 2019, \mn@doi [\mnras] {10.1093/mnras/sty2857}, \href
  {http://adsabs.harvard.edu/abs/2019MNRAS.483..370B} {483, 370}

\bibitem[\protect\citeauthoryear{{Bluck}, {Ellison}, {Patton}, {Simard},
  {Mendel}, {Teimoorinia}, {Moreno}  \& {Starkenburg}}{{Bluck}
  et~al.}{2014}]{Bluck2014}
{Bluck} A. F.~L.,  {Ellison} S.~L.,  {Patton} D.~R.,  {Simard} L.,  {Mendel}
  J.~T.,  {Teimoorinia} H.,  {Moreno} J.,   {Starkenburg} E.,  2014, arXiv
  e-prints, \href {https://ui.adsabs.harvard.edu/abs/2014arXiv1412.3862B} {p.
  arXiv:1412.3862}

\bibitem[\protect\citeauthoryear{{Bluck} et~al.,}{{Bluck}
  et~al.}{2016}]{Bluck2016}
{Bluck} A. F.~L.,  et~al., 2016, \mn@doi [\mnras] {10.1093/mnras/stw1665},
  \href {https://ui.adsabs.harvard.edu/abs/2016MNRAS.462.2559B} {462, 2559}

\bibitem[\protect\citeauthoryear{Bouch{\'{e}} et~al.,}{Bouch{\'{e}}
  et~al.}{2010}]{Bouche2010}
Bouch{\'{e}} N.,  et~al., 2010, \mn@doi [The Astrophysical Journal]
  {10.1088/0004-637X/718/2/1001}, 718, 1001

\bibitem[\protect\citeauthoryear{{Buta} et~al.,}{{Buta}
  et~al.}{2019}]{Buta2019}
{Buta} R.~J.,  et~al., 2019, \mn@doi [\mnras] {10.1093/mnras/stz1780}, \href
  {https://ui.adsabs.harvard.edu/abs/2019MNRAS.488.2175B} {488, 2175}

\bibitem[\protect\citeauthoryear{{Cicone} et~al.,}{{Cicone}
  et~al.}{2014}]{Cicone2014}
{Cicone} C.,  et~al., 2014, \mn@doi [\aap] {10.1051/0004-6361/201322464}, \href
  {http://adsabs.harvard.edu/abs/2014A%26A...562A..21C} {562, A21}

\bibitem[\protect\citeauthoryear{{Cluver} et~al.,}{{Cluver}
  et~al.}{2014}]{Cluver2014}
{Cluver} M.~E.,  et~al., 2014, \mn@doi [\apj] {10.1088/0004-637X/782/2/90},
  \href {https://ui.adsabs.harvard.edu/abs/2014ApJ...782...90C} {782, 90}

\bibitem[\protect\citeauthoryear{{Cluver}, {Jarrett}, {Dale}, {Smith}, {August}
   \& {Brown}}{{Cluver} et~al.}{2017}]{Cluver2017}
{Cluver} M.~E.,  {Jarrett} T.~H.,  {Dale} D.~A.,  {Smith} J. D.~T.,  {August}
  T.,   {Brown} M.~J.~I.,  2017, \mn@doi [\apj] {10.3847/1538-4357/aa92c7},
  \href {https://ui.adsabs.harvard.edu/abs/2017ApJ...850...68C} {850, 68}

\bibitem[\protect\citeauthoryear{{Cochrane} \& {Best}}{{Cochrane} \&
  {Best}}{2018}]{Cochrane2018}
{Cochrane} R.~K.,  {Best} P.~N.,  2018, \mn@doi [\mnras]
  {10.1093/mnras/sty1708}, \href
  {http://adsabs.harvard.edu/abs/2018MNRAS.480..864C} {480, 864}

\bibitem[\protect\citeauthoryear{{Cook}, {Cortese}, {Catinella}  \&
  {Robotham}}{{Cook} et~al.}{2019}]{Cook2019}
{Cook} R. H.~W.,  {Cortese} L.,  {Catinella} B.,   {Robotham} A.,  2019,
  \mn@doi [\mnras] {10.1093/mnras/stz2789}, \href
  {https://ui.adsabs.harvard.edu/abs/2019MNRAS.490.4060C} {490, 4060}

\bibitem[\protect\citeauthoryear{{Cortese}, {Catinella}, {Boissier}, {Boselli}
  \& {Heinis}}{{Cortese} et~al.}{2011}]{Cortese2011}
{Cortese} L.,  {Catinella} B.,  {Boissier} S.,  {Boselli} A.,   {Heinis} S.,
  2011, \mn@doi [\mnras] {10.1111/j.1365-2966.2011.18822.x}, \href
  {https://ui.adsabs.harvard.edu/abs/2011MNRAS.415.1797C} {415, 1797}

\bibitem[\protect\citeauthoryear{{D{\'e}nes}, {Kilborn}  \&
  {Koribalski}}{{D{\'e}nes} et~al.}{2014}]{Denes2014}
{D{\'e}nes} H.,  {Kilborn} V.~A.,   {Koribalski} B.~S.,  2014, \mn@doi [\mnras]
  {10.1093/mnras/stu1337}, \href
  {https://ui.adsabs.harvard.edu/abs/2014MNRAS.444..667D} {444, 667}

\bibitem[\protect\citeauthoryear{{Durbala}, {Sulentic}, {Buta}  \&
  {Verdes-Montenegro}}{{Durbala} et~al.}{2008}]{Durbala2008}
{Durbala} A.,  {Sulentic} J.~W.,  {Buta} R.,   {Verdes-Montenegro} L.,  2008,
  \mn@doi [\mnras] {10.1111/j.1365-2966.2008.13713.x}, \href
  {https://ui.adsabs.harvard.edu/abs/2008MNRAS.390..881D} {390, 881}

\bibitem[\protect\citeauthoryear{{Elbaz} et~al.,}{{Elbaz}
  et~al.}{2011}]{Elbaz2011}
{Elbaz} D.,  et~al., 2011, \mn@doi [\aap] {10.1051/0004-6361/201117239}, \href
  {https://ui.adsabs.harvard.edu/abs/2011A&A...533A.119E} {533, A119}

\bibitem[\protect\citeauthoryear{{Ellison}, {Patton}, {Simard}, {McConnachie},
  {Baldry}  \& {Mendel}}{{Ellison} et~al.}{2010}]{Ellison2010}
{Ellison} S.~L.,  {Patton} D.~R.,  {Simard} L.,  {McConnachie} A.~W.,  {Baldry}
  I.~K.,   {Mendel} J.~T.,  2010, \mn@doi [\mnras]
  {10.1111/j.1365-2966.2010.17076.x}, \href
  {http://adsabs.harvard.edu/abs/2010MNRAS.407.1514E} {407, 1514}

\bibitem[\protect\citeauthoryear{{Espada}, {Verdes-Montenegro}, {Huchtmeier},
  {Sulentic}, {Verley}, {Leon}  \& {Sabater}}{{Espada}
  et~al.}{2011}]{Espada2011}
{Espada} D.,  {Verdes-Montenegro} L.,  {Huchtmeier} W.~K.,  {Sulentic} J.,
  {Verley} S.,  {Leon} S.,   {Sabater} J.,  2011, \mn@doi [\aap]
  {10.1051/0004-6361/201016117}, \href
  {https://ui.adsabs.harvard.edu/abs/2011A&A...532A.117E} {532, A117}

\bibitem[\protect\citeauthoryear{{Fern{\'a}ndez Lorenzo}, {Sulentic},
  {Verdes-Montenegro}  \& {Argudo-Fern{\'a}ndez}}{{Fern{\'a}ndez Lorenzo}
  et~al.}{2013}]{FernandezLorenzo2013}
{Fern{\'a}ndez Lorenzo} M.,  {Sulentic} J.,  {Verdes-Montenegro} L.,
  {Argudo-Fern{\'a}ndez} M.,  2013, \mn@doi [\mnras] {10.1093/mnras/stt1020},
  \href {http://adsabs.harvard.edu/abs/2013MNRAS.434..325F} {434, 325}

\bibitem[\protect\citeauthoryear{{Giavalisco} et~al.,}{{Giavalisco}
  et~al.}{2004}]{Giavalisco2004}
{Giavalisco} M.,  et~al., 2004, \mn@doi [\apjl] {10.1086/379232}, \href
  {http://adsabs.harvard.edu/abs/2004ApJ...600L..93G} {600, L93}

\bibitem[\protect\citeauthoryear{{Giovanelli} et~al.,}{{Giovanelli}
  et~al.}{2005}]{Giovanelli2005}
{Giovanelli} R.,  et~al., 2005, \mn@doi [\aj] {10.1086/497431}, \href
  {http://adsabs.harvard.edu/abs/2005AJ....130.2598G} {130, 2598}

\bibitem[\protect\citeauthoryear{Gunn \& {Gott, J. Richard}}{Gunn \& {Gott, J.
  Richard}}{1972}]{Gunn1972}
Gunn J.~E.,  {Gott, J. Richard} I.,  1972, \mn@doi [The Astrophysical Journal]
  {10.1086/151605}, 176, 1

\bibitem[\protect\citeauthoryear{{Haynes} \& {Giovanelli}}{{Haynes} \&
  {Giovanelli}}{1984}]{Haynes1984}
{Haynes} M.~P.,  {Giovanelli} R.,  1984, \mn@doi [\aj] {10.1086/113573}, \href
  {https://ui.adsabs.harvard.edu/abs/1984AJ.....89..758H} {89, 758}

\bibitem[\protect\citeauthoryear{{Haynes} et~al.,}{{Haynes}
  et~al.}{2011}]{Haynes2011}
{Haynes} M.~P.,  et~al., 2011, \mn@doi [\aj] {10.1088/0004-6256/142/5/170},
  \href {http://adsabs.harvard.edu/abs/2011AJ....142..170H} {142, 170}

\bibitem[\protect\citeauthoryear{{Haynes} et~al.,}{{Haynes}
  et~al.}{2018}]{Haynes2018}
{Haynes} M.~P.,  et~al., 2018, \mn@doi [\apj] {10.3847/1538-4357/aac956}, \href
  {https://ui.adsabs.harvard.edu/abs/2018ApJ...861...49H} {861, 49}

\bibitem[\protect\citeauthoryear{{Jarrett}, {Helou}, {Van Buren}, {Valjavec}
  \& {Condon}}{{Jarrett} et~al.}{1999}]{Jarrett1999}
{Jarrett} T.~H.,  {Helou} G.,  {Van Buren} D.,  {Valjavec} E.,   {Condon}
  J.~J.,  1999, \mn@doi [\aj] {10.1086/301080}, \href
  {https://ui.adsabs.harvard.edu/abs/1999AJ....118.2132J} {118, 2132}

\bibitem[\protect\citeauthoryear{{Jarrett} et~al.,}{{Jarrett}
  et~al.}{2011}]{Jarrett2011}
{Jarrett} T.~H.,  et~al., 2011, \mn@doi [\apj] {10.1088/0004-637X/735/2/112},
  \href {https://ui.adsabs.harvard.edu/abs/2011ApJ...735..112J} {735, 112}

\bibitem[\protect\citeauthoryear{{Jarrett} et~al.,}{{Jarrett}
  et~al.}{2012}]{Jarrett2012}
{Jarrett} T.~H.,  et~al., 2012, \mn@doi [\aj] {10.1088/0004-6256/144/2/68},
  \href {https://ui.adsabs.harvard.edu/abs/2012AJ....144...68J} {144, 68}

\bibitem[\protect\citeauthoryear{{Jarrett} et~al.,}{{Jarrett}
  et~al.}{2013}]{Jarrett2013}
{Jarrett} T.~H.,  et~al., 2013, \mn@doi [\aj] {10.1088/0004-6256/145/1/6},
  \href {https://ui.adsabs.harvard.edu/abs/2013AJ....145....6J} {145, 6}

\bibitem[\protect\citeauthoryear{{Jarrett}, {Cluver}, {Brown}, {Dale}, {Tsai}
  \& {Masci}}{{Jarrett} et~al.}{2019}]{Jarrett2019}
{Jarrett} T.~H.,  {Cluver} M.~E.,  {Brown} M.~J.~I.,  {Dale} D.~A.,  {Tsai}
  C.~W.,   {Masci} F.,  2019, \mn@doi [\apjs] {10.3847/1538-4365/ab521a}, \href
  {https://ui.adsabs.harvard.edu/abs/2019ApJS..245...25J} {245, 25}

\bibitem[\protect\citeauthoryear{{Jones} et~al.,}{{Jones}
  et~al.}{2018}]{Jones2018}
{Jones} M.~G.,  et~al., 2018, \mn@doi [\aap] {10.1051/0004-6361/201731448},
  \href {https://ui.adsabs.harvard.edu/abs/2018A%26A...609A..17J} {609, A17}

\bibitem[\protect\citeauthoryear{Karachentseva}{Karachentseva}{1973}]{Karachentseva1973}
Karachentseva V.~E.,  1973, Astrof. Issledovanija Byu. Spec. Ast. Obs.; Vol. 8;
  Page 3-49, 8

\bibitem[\protect\citeauthoryear{{Kormendy} \& {Kennicutt}}{{Kormendy} \&
  {Kennicutt}}{2004}]{KormendyKennicutt2004}
{Kormendy} J.,  {Kennicutt} Robert~C. J.,  2004, \mn@doi [\araa]
  {10.1146/annurev.astro.42.053102.134024}, \href
  {https://ui.adsabs.harvard.edu/abs/2004ARA&A..42..603K} {42, 603}

\bibitem[\protect\citeauthoryear{Lee et~al.,}{Lee et~al.}{2015}]{Lee2015}
Lee N.,  et~al., 2015, \mn@doi [The Astrophysical Journal]
  {10.1088/0004-637X/801/2/80}, 801, 80

\bibitem[\protect\citeauthoryear{{Leon} et~al.,}{{Leon}
  et~al.}{2008}]{Leon2008}
{Leon} S.,  et~al., 2008, \mn@doi [\aap] {10.1051/0004-6361:20078533}, \href
  {https://ui.adsabs.harvard.edu/abs/2008A&A...485..475L} {485, 475}

\bibitem[\protect\citeauthoryear{{Lisenfeld} et~al.,}{{Lisenfeld}
  et~al.}{2007}]{Lisenfeld2007}
{Lisenfeld} U.,  et~al., 2007, in {Combes} F.,  {Palou{\v{s}}} J.,  eds,  IAU
  Symposium Vol. 235, Galaxy Evolution across the Hubble Time. pp 219--219,
  \mn@doi{10.1017/S1743921306006259}

\bibitem[\protect\citeauthoryear{{Liu} \& {Cen}}{{Liu} \&
  {Cen}}{2017}]{Liu2017}
{Liu} J.,  {Cen} R.,  2017, arXiv e-prints, \href
  {https://ui.adsabs.harvard.edu/abs/2017arXiv170100866L} {p. arXiv:1701.00866}

\bibitem[\protect\citeauthoryear{{Martig}, {Bournaud}, {Teyssier}  \&
  {Dekel}}{{Martig} et~al.}{2009}]{Martig2009}
{Martig} M.,  {Bournaud} F.,  {Teyssier} R.,   {Dekel} A.,  2009, \mn@doi
  [\apj] {10.1088/0004-637X/707/1/250}, \href
  {http://adsabs.harvard.edu/abs/2009ApJ...707..250M} {707, 250}

\bibitem[\protect\citeauthoryear{{McNamara} \& {Nulsen}}{{McNamara} \&
  {Nulsen}}{2007}]{McNamara&Nulsen2007}
{McNamara} B.~R.,  {Nulsen} P.~E.~J.,  2007, \mn@doi [\araa]
  {10.1146/annurev.astro.45.051806.110625}, \href
  {http://adsabs.harvard.edu/abs/2007ARA%26A..45..117M} {45, 117}

\bibitem[\protect\citeauthoryear{{McNamara}, {Wise}, {David}, {Nulsen}  \&
  {Sarazin}}{{McNamara} et~al.}{2000}]{McNamara2000}
{McNamara} B.~R.,  {Wise} M.~W.,  {David} L.~P.,  {Nulsen} P.~E.~J.,
  {Sarazin} C.~L.,  2000, in AAS/High Energy Astrophysics Division \#5. p.~1201

\bibitem[\protect\citeauthoryear{{McPartland}, {Sanders}, {Kewley}  \&
  {Leslie}}{{McPartland} et~al.}{2018}]{McPartland2018}
{McPartland} C.,  {Sanders} D.~B.,  {Kewley} L.~J.,   {Leslie} S.~K.,  2018,
  \mn@doi [\mnras] {10.1093/mnrasl/sly202}, \href
  {https://arxiv.org/pdf/1810.10021.pdf} {482, 129}

\bibitem[\protect\citeauthoryear{{Moon}, {An}  \& {Yoon}}{{Moon}
  et~al.}{2019}]{Moon2019}
{Moon} J.-S.,  {An} S.-H.,   {Yoon} S.-J.,  2019, \mn@doi [\apj]
  {10.3847/1538-4357/ab3401}, \href
  {https://ui.adsabs.harvard.edu/abs/2019ApJ...882...14M} {882, 14}

\bibitem[\protect\citeauthoryear{Moore, Katz, Lake, Dressler  \& Oemler}{Moore
  et~al.}{1995}]{Moore1996}
Moore B.,  Katz N.,  Lake G.,  Dressler A.,   Oemler A.,  1995, \mn@doi
  [Nature, Volume 379, Issue 6566, pp. 613-616 (1996).] {10.1038/379613a0},
  379, 613

\bibitem[\protect\citeauthoryear{{Morselli}, {Popesso}, {Cibinel}, {Oesch},
  {Montes}, {Atek}, {Illingworth}  \& {Holden}}{{Morselli}
  et~al.}{2018}]{Morselli2018}
{Morselli} L.,  {Popesso} P.,  {Cibinel} A.,  {Oesch} P.~A.,  {Montes} M.,
  {Atek} H.,  {Illingworth} G.~D.,   {Holden} B.,  2018, arXiv e-prints, \href
  {http://adsabs.harvard.edu/abs/2018arXiv181208561M} {}

\bibitem[\protect\citeauthoryear{Noeske et~al.,}{Noeske
  et~al.}{2007}]{Noeske2007}
Noeske K.~G.,  et~al., 2007, \mn@doi [The Astrophysical Journal]
  {10.1086/517926}, 660, L43

\bibitem[\protect\citeauthoryear{{Nulsen}}{{Nulsen}}{1982}]{Nulsen1982}
{Nulsen} P.~E.~J.,  1982, \mn@doi [\mnras] {10.1093/mnras/198.4.1007}, \href
  {https://ui.adsabs.harvard.edu/abs/1982MNRAS.198.1007N} {198, 1007}

\bibitem[\protect\citeauthoryear{{Parkash}, {Brown}, {Jarrett}  \&
  {Bonne}}{{Parkash} et~al.}{2018}]{Parkash2018}
{Parkash} V.,  {Brown} M.~J.~I.,  {Jarrett} T.~H.,   {Bonne} N.~J.,  2018,
  \mn@doi [\apj] {10.3847/1538-4357/aad3b9}, \href
  {http://adsabs.harvard.edu/abs/2018ApJ...864...40P} {864, 40}

\bibitem[\protect\citeauthoryear{{Pearson} et~al.,}{{Pearson}
  et~al.}{2019}]{Pearson2019}
{Pearson} W.~J.,  et~al., 2019, arXiv e-prints, \href
  {https://ui.adsabs.harvard.edu/abs/2019arXiv190810115P} {p. arXiv:1908.10115}

\bibitem[\protect\citeauthoryear{Peng, Maiolino  \& Cochrane}{Peng
  et~al.}{2015}]{Peng2015}
Peng Y.,  Maiolino R.,   Cochrane R.,  2015, \mn@doi [Nature, Volume 521, Issue
  7551, pp. 192-195 (2015).] {10.1038/nature14439}, 521, 192

\bibitem[\protect\citeauthoryear{{Robotham} et~al.,}{{Robotham}
  et~al.}{2014}]{Robotham2014}
{Robotham} A.~S.~G.,  et~al., 2014, \mn@doi [\mnras] {10.1093/mnras/stu1604},
  \href {http://adsabs.harvard.edu/abs/2014MNRAS.444.3986R} {444, 3986}

\bibitem[\protect\citeauthoryear{{Rodighiero} et~al.,}{{Rodighiero}
  et~al.}{2011}]{Rodighiero2011}
{Rodighiero} G.,  et~al., 2011, \mn@doi [\apjl] {10.1088/2041-8205/739/2/L40},
  \href {https://ui.adsabs.harvard.edu/abs/2011ApJ...739L..40R} {739, L40}

\bibitem[\protect\citeauthoryear{{Sabater}, {Leon}, {Verdes-Montenegro},
  {Lisenfeld}, {Sulentic}  \& {Verley}}{{Sabater} et~al.}{2008}]{Sabater2008}
{Sabater} J.,  {Leon} S.,  {Verdes-Montenegro} L.,  {Lisenfeld} U.,  {Sulentic}
  J.,   {Verley} S.,  2008, \mn@doi [\aap] {10.1051/0004-6361:20078785}, \href
  {https://ui.adsabs.harvard.edu/abs/2008A&A...486...73S} {486, 73}

\bibitem[\protect\citeauthoryear{{Sabater}, {Verdes-Montenegro}, {Leon},
  {Sulentic}, {Lisenfeld}  \& {Verley}}{{Sabater} et~al.}{2011}]{Sabater2011}
{Sabater} J.,  {Verdes-Montenegro} L.,  {Leon} S.,  {Sulentic} J.,  {Lisenfeld}
  U.,   {Verley} S.,  2011, in {Zapatero Osorio} M.~R.,  {Gorgas} J.,
  {Ma{\'\i}z Apell{\'a}niz} J.,  {Pardo} J.~R.,   {Gil de Paz} A.,  eds,
  Highlights of Spanish Astrophysics VI. pp 399--399

\bibitem[\protect\citeauthoryear{{Saintonge} et~al.,}{{Saintonge}
  et~al.}{2016}]{Saintonge2016}
{Saintonge} A.,  et~al., 2016, \mn@doi [\mnras] {10.1093/mnras/stw1715}, \href
  {https://ui.adsabs.harvard.edu/abs/2016MNRAS.462.1749S} {462, 1749}

\bibitem[\protect\citeauthoryear{Scholz \& Stephens}{Scholz \&
  Stephens}{1987}]{Scholz1987}
Scholz F.~W.,  Stephens M.~A.,  1987, \mn@doi [Journal of the American
  Statistical Association] {10.1080/01621459.1987.10478517}, 82, 918

\bibitem[\protect\citeauthoryear{{Schreiber} et~al.,}{{Schreiber}
  et~al.}{2015}]{Schreiber2015}
{Schreiber} C.,  et~al., 2015, \mn@doi [\aap] {10.1051/0004-6361/201425017},
  \href {https://ui.adsabs.harvard.edu/abs/2015A&A...575A..74S} {575, A74}

\bibitem[\protect\citeauthoryear{{Speagle}, {Steinhardt}, {Capak}  \&
  {Silverman}}{{Speagle} et~al.}{2014}]{Speagle2014}
{Speagle} J.~S.,  {Steinhardt} C.~L.,  {Capak} P.~L.,   {Silverman} J.~D.,
  2014, \mn@doi [\apjs] {10.1088/0067-0049/214/2/15}, \href
  {https://ui.adsabs.harvard.edu/abs/2014ApJS..214...15S} {214, 15}

\bibitem[\protect\citeauthoryear{{Sulentic} et~al.,}{{Sulentic}
  et~al.}{2006}]{Sulentic2006}
{Sulentic} J.~W.,  et~al., 2006, \mn@doi [\aap] {10.1051/0004-6361:20054020},
  \href {https://ui.adsabs.harvard.edu/abs/2006A&A...449..937S} {449, 937}

\bibitem[\protect\citeauthoryear{{Tacconi} et~al.,}{{Tacconi}
  et~al.}{2013}]{Tacconi2013}
{Tacconi} L.~J.,  et~al., 2013, \mn@doi [\apj] {10.1088/0004-637X/768/1/74},
  \href {http://adsabs.harvard.edu/abs/2013ApJ...768...74T} {768, 74}

\bibitem[\protect\citeauthoryear{{Tacconi} et~al.,}{{Tacconi}
  et~al.}{2018}]{Tacconi2018}
{Tacconi} L.~J.,  et~al., 2018, \mn@doi [\apj] {10.3847/1538-4357/aaa4b4},
  \href {http://adsabs.harvard.edu/abs/2018ApJ...853..179T} {853, 179}

\bibitem[\protect\citeauthoryear{Verdes-Montenegro, Sulentic, Lisenfeld, Leon,
  Espada, Garcia, Sabater  \& Verley}{Verdes-Montenegro
  et~al.}{2005}]{VerdesMontenegro2005}
Verdes-Montenegro L.,  Sulentic J.,  Lisenfeld U.,  Leon S.,  Espada D.,
  Garcia E.,  Sabater J.,   Verley S.,  2005, \mn@doi [Astronomy and
  Astrophysics, Volume 436, Issue 2, June III 2005, pp.443-455]
  {10.1051/0004-6361:20042280}, 436, 443

\bibitem[\protect\citeauthoryear{Verley et~al.,}{Verley
  et~al.}{2007a}]{Verley2007}
Verley S.,  et~al., 2007a, \mn@doi [Astronomy and Astrophysics]
  {10.1051/0004-6361:20077307}, 470, 505

\bibitem[\protect\citeauthoryear{Verley et~al.,}{Verley
  et~al.}{2007b}]{Verley2007a}
Verley S.,  et~al., 2007b, \mn@doi [Astronomy and Astrophysics]
  {10.1051/0004-6361:20077481}, 472, 121

\bibitem[\protect\citeauthoryear{{Wang}, {Kong}  \& {Pan}}{{Wang}
  et~al.}{2018}]{Wang2018}
{Wang} E.,  {Kong} X.,   {Pan} Z.,  2018, \mn@doi [\apj]
  {10.3847/1538-4357/aadb9e}, \href
  {http://adsabs.harvard.edu/abs/2018ApJ...865...49W} {865, 49}

\bibitem[\protect\citeauthoryear{Whitaker et~al.,}{Whitaker
  et~al.}{2015}]{Whitaker2015}
Whitaker K.~E.,  et~al., 2015, \mn@doi [The Astrophysical Journal]
  {10.1088/2041-8205/811/1/L12}, 811, L12

\bibitem[\protect\citeauthoryear{{Wright} et~al.,}{{Wright}
  et~al.}{2010}]{Wright2010}
{Wright} E.~L.,  et~al., 2010, \mn@doi [\aj] {10.1088/0004-6256/140/6/1868},
  \href {https://ui.adsabs.harvard.edu/abs/2010AJ....140.1868W} {140, 1868}

\bibitem[\protect\citeauthoryear{{Xu} et~al.,}{{Xu} et~al.}{2010}]{Xu2010}
{Xu} C.~K.,  et~al., 2010, in American Astronomical Society Meeting Abstracts
  \#216. p. 307.03

\bibitem[\protect\citeauthoryear{{Xu} et~al.,}{{Xu} et~al.}{2012}]{Xu2012}
{Xu} C.~K.,  et~al., 2012, \mn@doi [\apj] {10.1088/0004-637X/760/1/72}, \href
  {https://ui.adsabs.harvard.edu/abs/2012ApJ...760...72X} {760, 72}

\makeatother
\end{thebibliography}

\section*{Appendix A: \textit{WISE} De-Blending of Resolved Galaxies}

Blending of resolved galaxies with other nearby (in projection) galaxies is a general problem for \textit{WISE} because
of its relatively large beam size, but is notably a feature of dense structures, such as galaxy clusters, compact groups
and pairs.   Accordingly, de-blending is a necessary component of \textit{WISE} galaxy characterization pipelines.  \cite{Jarrett2013,Jarrett2019} created a general processing pipeline that also includes expert interaction to carry out de-blending
operations.    The success of the de-blend for pair-galaxies depends on a few factors, including how close the pair members are located to each other,
the relative brightness, the relative orientation (e.g., disk orientation for spirals), and the asymmetry of one or
both of the pair galaxies.  Sources that are too close
(within or comparable to the beam of \textit{WISE}) may not have acceptable de-blended fluxes.   Sources that are faint
compared to their companion and located close to the companion core, may also not de-blend well.  Asymmetric
sources (e.g., significant tidal distortion) are not easily modeled.

The principle assumption with these algorithms is that the galaxy light can be modeled using axi-symmetric
averaging to produce a smoothed, underlying surface brightness distribution.  Each galaxy is modeled accordingly,
masking the nearby galaxy, measuring the light shape at the 3-sigma isophotal level, fitting an ellipsoid to the 2D image, and 
then the approximate ellipsoid shape is then used to 
 to subtract from the image instead of masking.  Each iteration of the process, moving from
one galaxy component to the next, creates a model that is closer to the actual light distribution for each galaxy of the
pair.   Several iterations are executed until convergence of the fitting metric is achieved.   At this point, the expert user
inspects the result, and makes adjustments to the original fitting parameters, including the disk orientation and axis 
ratio (for each galaxy in the pair),   as well any other emission that may not be relevant to the galaxy pair (stars, 
background galaxies, artifacts).   The user then runs the pipeline with these new parameters, and repeats the cycle
as needed.   Additional masking may be required for residual subtractions near the nuclei and other high 
surface brightness regions.
 The de-blend tends to work best for those pairs that have more than half their light unaffected by the
blend (i.e., they are well separated).   It becomes a significant challenge when most of the light is over-lapping
(e.g., with physical merging systems) or when the disk orientations are parallel (disks overlap in the same direction).
A relatively simple case is for a blue-red pair, where one galaxy is star-forming, and the other is passive or quenched
(appearing blue in \textit{WISE} colors), because only two bands -- W1 [3.4$\, \mu$m]  and W2 [4.6$\, \mu$m] -- have any blending.  
Whereas, for two
SF galaxies all four \textit{WISE} bands require de-blend procedures.

\begin{figure*}

\centering
 \includegraphics[width=160mm]{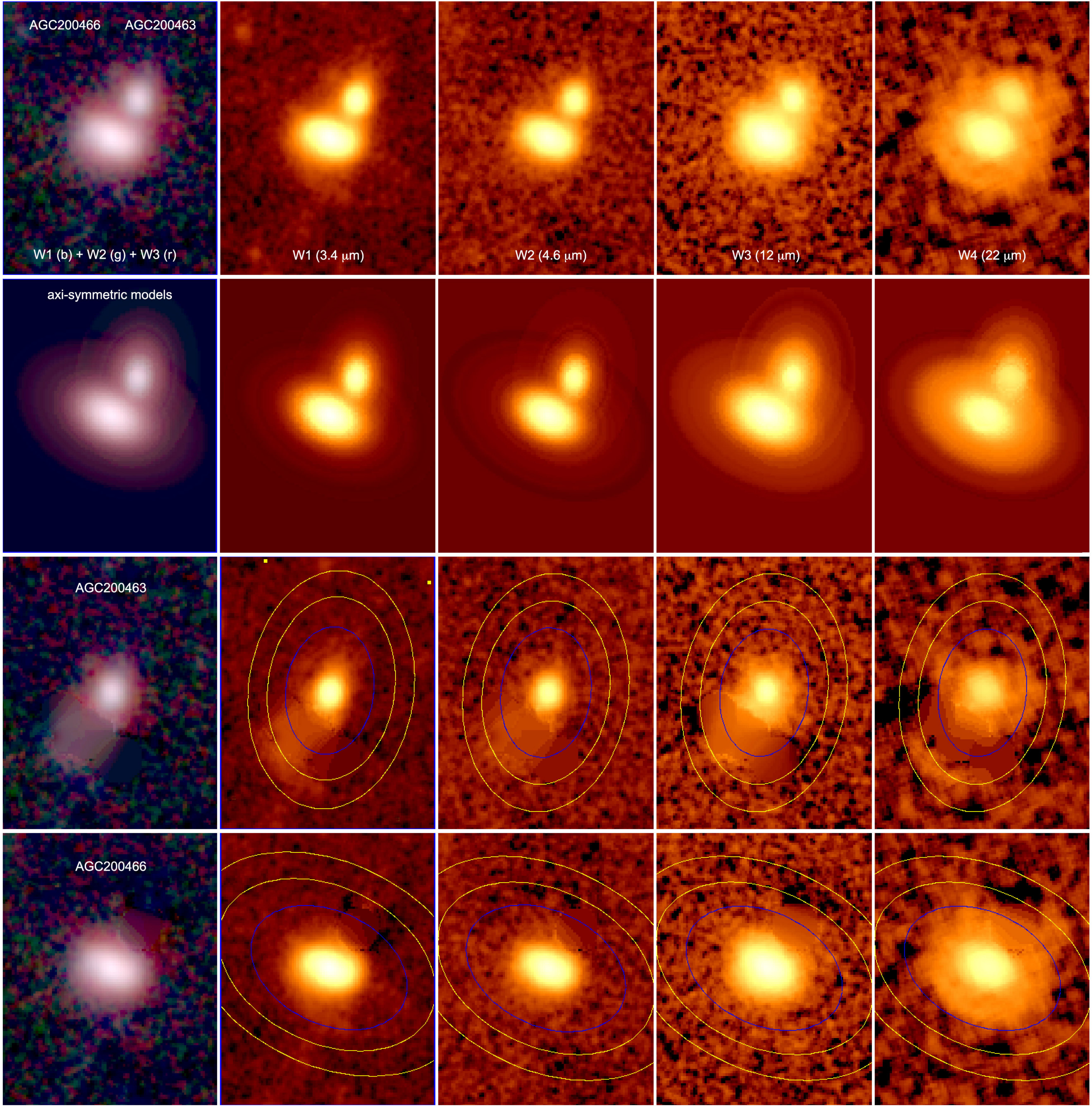}
\caption{\textit{WISE} view of the galaxy pair AGC200466 and AGC200463, de-blended using the \textit{WISE} user-interaction pipeline.  The top panel shows the four bands of \textit{WISE}, and a 3-color combination.  Note both galaxies are gas-rich and star-forming; hence all four bands require de-blend.  The middle panel shows the resulting axi-symmetric models that represent the smoothed and symmetric emission from each galaxy.  The bottom two panels show the resulting de-blend for each galaxy pair, with their respective 1-$\sigma$ isophotal apertures (blue ellipse) and background sky annuli (yellow).  The field-of-view is 46 $\arcsec$.}
\label{fig:deblend}
\end{figure*}

An example of a galaxy pair in this current study that is clearly blended is the AGC\,200466 -- AGC\,200463 system,
both of which are star-forming galaxies.  The nuclei are separated by 20\arcsec (12.92 kpc projection) and approximately half of their
light is overlapping with each other.   Fig.~\ref{fig:deblend} shows the four bands of \textit{WISE} for this galaxy pair.
It is clearly seen that AGC\,200466 is the brighter of the two, and hence we can expect its final de-blend to
be higher quality than its fainter companion (200463).   The de-blend pipeline, after iterating the masking and 
subtraction procedure, constructs the axi-symmetric models, shown in the second-row panel of Fig.~\ref{fig:deblend}.
Visually the models appear highly satisfactory, notably for the short-wavelength bands where the surface brightness is higher
and angular resolution is better.    \cite{Jarrett2012} note that the \textit{WISE} angular resolution for drizzle-mosaiced images
is about 6.0, 6.5, 7.1 and 12.4\arcsec for W1 [3.4$\, \mu$m], W2 [4.6$\, \mu$m], W3 [12$\, \mu$m]  and W4 [22$\, \mu$m],
respectively.  The de-blend pipeline assumes the same axi-symmetric shape (based on W1 measurements) for
all four bands, which is simplistic for the longer wavelengths, but generally adequate for de-blend purposes.  Some additional
circular, small-radius masking was carried out to suppress large subtraction residuals.  The resulting de-blended image
for each galaxy of the pair is seen in the bottom panels of Fig.~\ref{fig:deblend}.  The brighter companion (200466) has
an excellent solution in all four bands: the de-blend was successful.   The fainter companion (200463) appears to be
acceptable, although it clearly has some residual emission from the brighter companion to the south-east.   This extra
emission is estimated to be less than 10\% of the total, for this case.

\begin{figure*}

\centering
 \includegraphics[width=70mm]{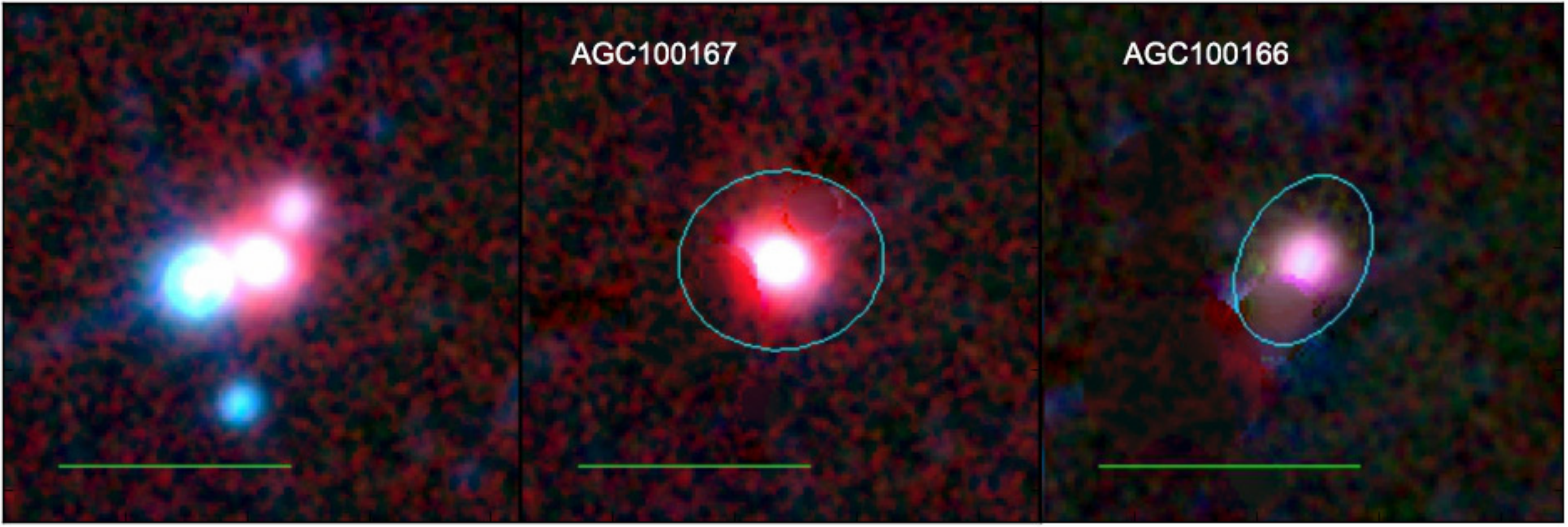}
  \includegraphics[width=70mm]{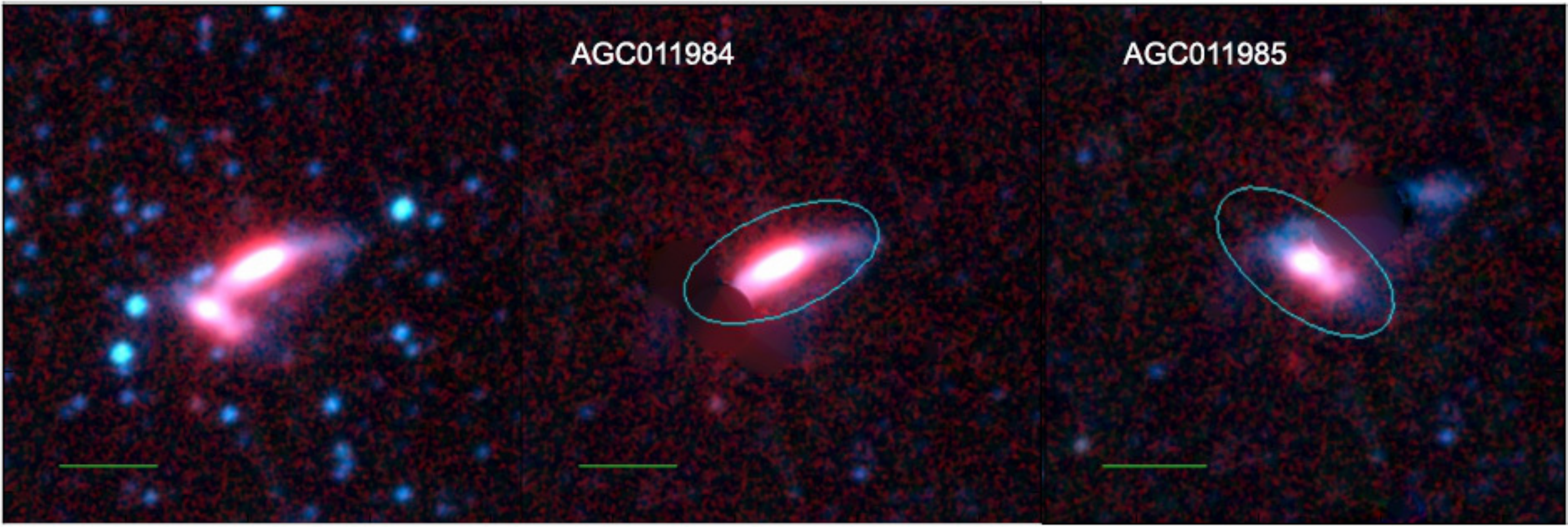}
   \includegraphics[width=70mm]{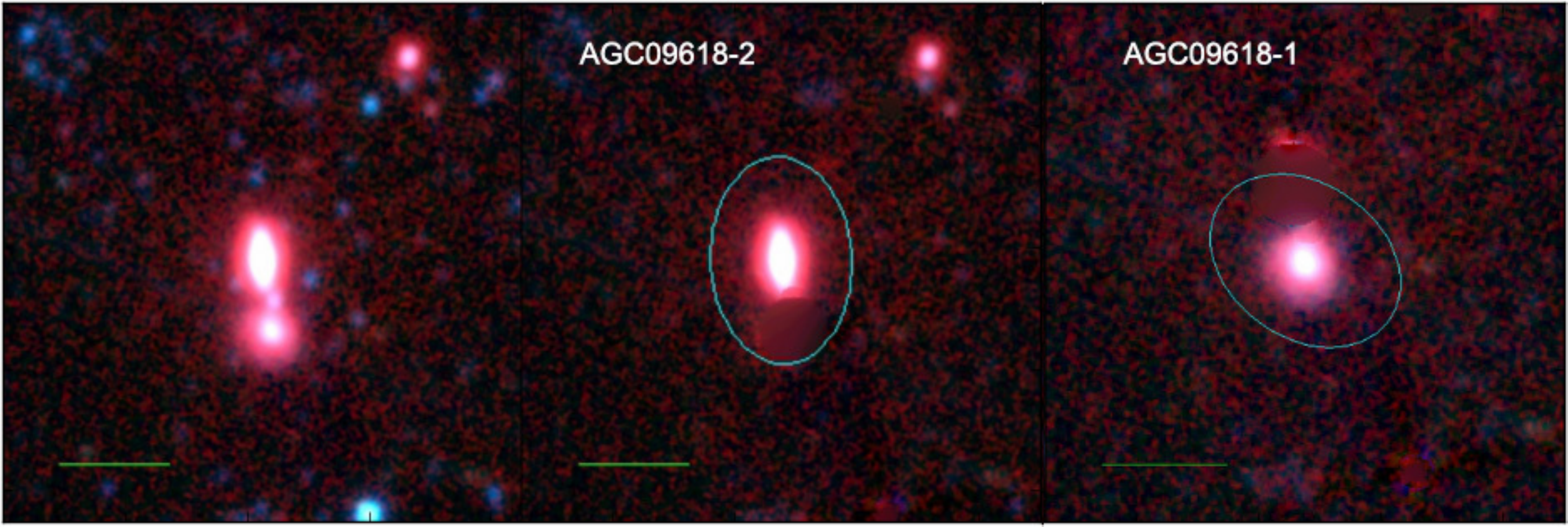}
    \includegraphics[width=70mm]{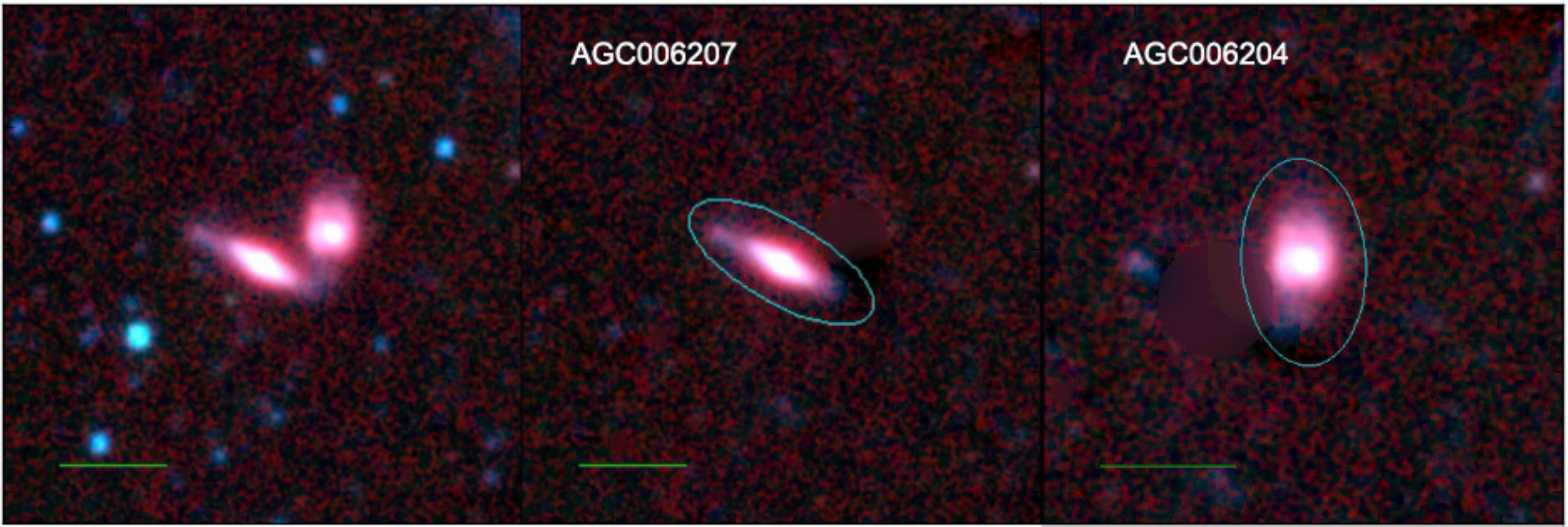}
   \includegraphics[width=70mm]{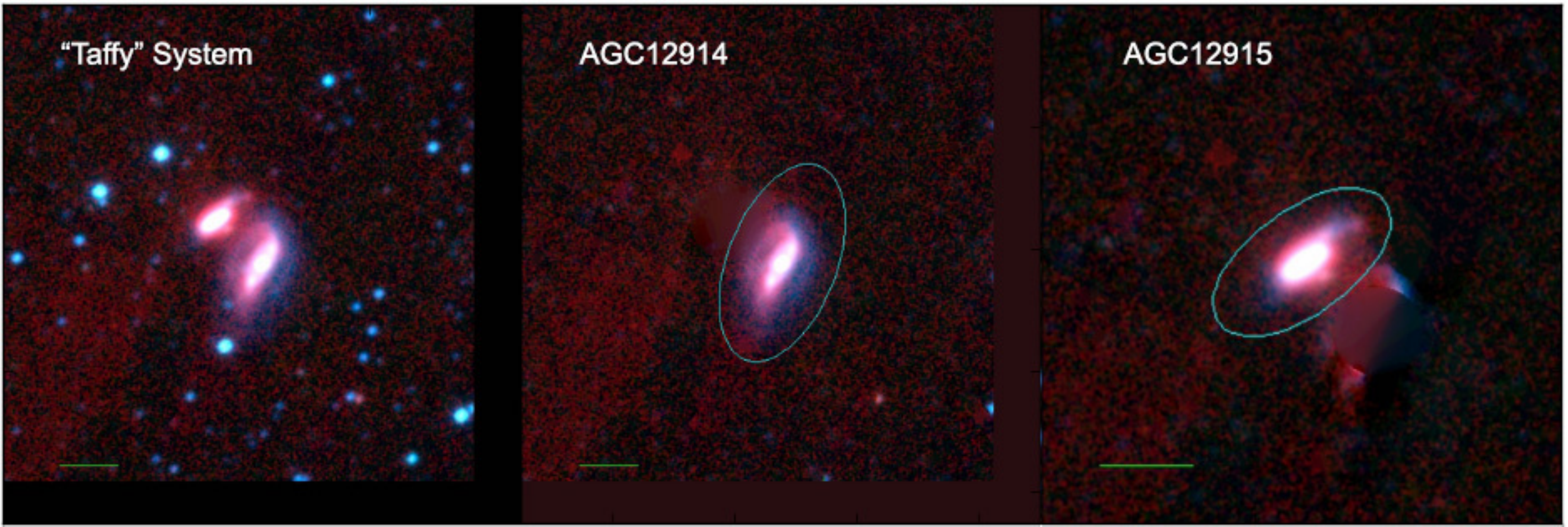}

 \caption{De-blended pair systems in this study.   These five systems have been identified has having significant blending, notably
the first three in the sequence shown here.  The images are visualized using three bands of \textit{WISE}:  W1 [3.4$\, \mu$m]  in blue,
W2 [4.6$\, \mu$m] in green, and W3 [12$\, \mu$m] in red.  The green dash indicates 1 arcmin scale.
\label{fig:deblends}}
\end{figure*}

\begin{figure*}
\includegraphics[scale = 0.5]{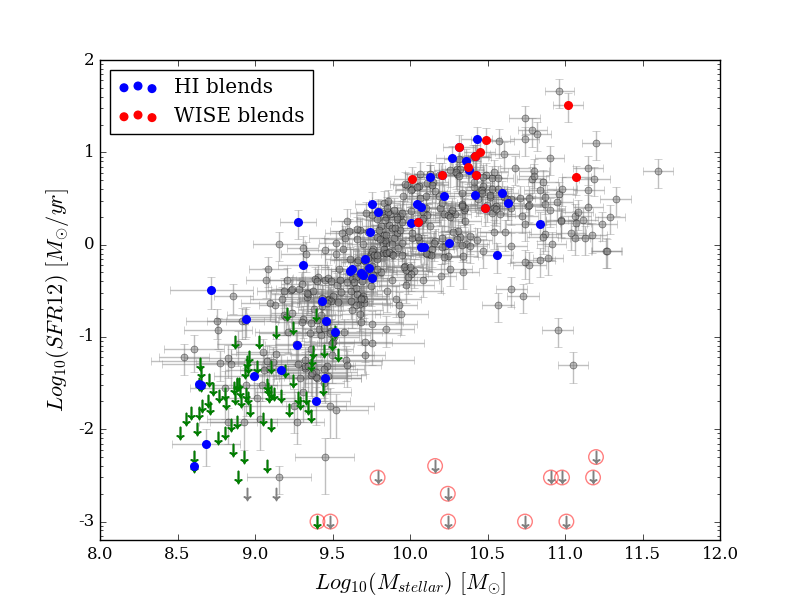}
\caption{The pair MS of Figure \ref{MS}b with the location of all noted HI blends highlighted in blue, and \textit{WISE} blends highlighted in red. Based on the locations of these blends (i.e. not outliers) we pose that blending in general does not appear to be driving the scatter in the pair MS, nor the consequent observed difference in scatter between the pair and isolated galaxy samples.}
\label{Allblends}
\end{figure*}

In addition to the AGC\,20466/3 system, we identify another five pairs in this study that have de-blended measurements.
We provide the 3-color (W1 + W2 + W3) images of these systems, along with their solutions (again in 3-color),
shown in Fig~\ref{fig:deblends}.   All of these pairs are star-forming, and are relatively high stellar mass systems
(see Figure \ref{Allblends}).  The de-blend solutions are adequate, with 10-20\%
additional uncertainty, but well within the scatter of the SF-MS.

The first example, AGC100166/7 shows a bright primary and a relatively faint secondary, while also contaminated by
a nearby star.   The star is 'blue',  meaning it is only bright in W1 and W2, so it has potential to alter the stellar mass 
estimation, but not the star formation.  The de-blend solution is satisfactory for the primary (100167), but may have some
excess W3 emission for the secondary (100166).   The second example is AGC\,011984/5,  two inclined disks blending
at a sharp angle, with the brighter companion (11984) having its outer disk cross the nucleus of the fainter companion
(11985).  This represents a challenging de-blend case because of the crossing orientation. It is clear that the brighter
companion has its outer disk  reduced by the de-blend and recovery of the secondary (whose resulting de-blend looks
good).   The third example, AGC09618-1/2,  is an insidious case in which the two disks are blending in parallel, making
it difficult to distinguish between the two.  The de-blend solutions look adequate to the eye, but clearly there is some
uncertainty in this case.   The final two cases have pair angular separations such that the de-blend solution was clean,
and clearly extracted sources successfully.  It is interesting to note, the last example, AGC12914/5, is the famous 
"Taffy" interaction system, in which both galaxies are greatly disturbed -- and hence highly asymmetric -- and likely
active by tidal-triggering \citep{Jarrett1999}.  Nevertheless, the separations are large enough for the de-blend pipeline to clearly separate the 
pair components.

\section*{Appendix B: Example cases of blending in HI}
\begin{figure*}
\includegraphics[width=7cm]{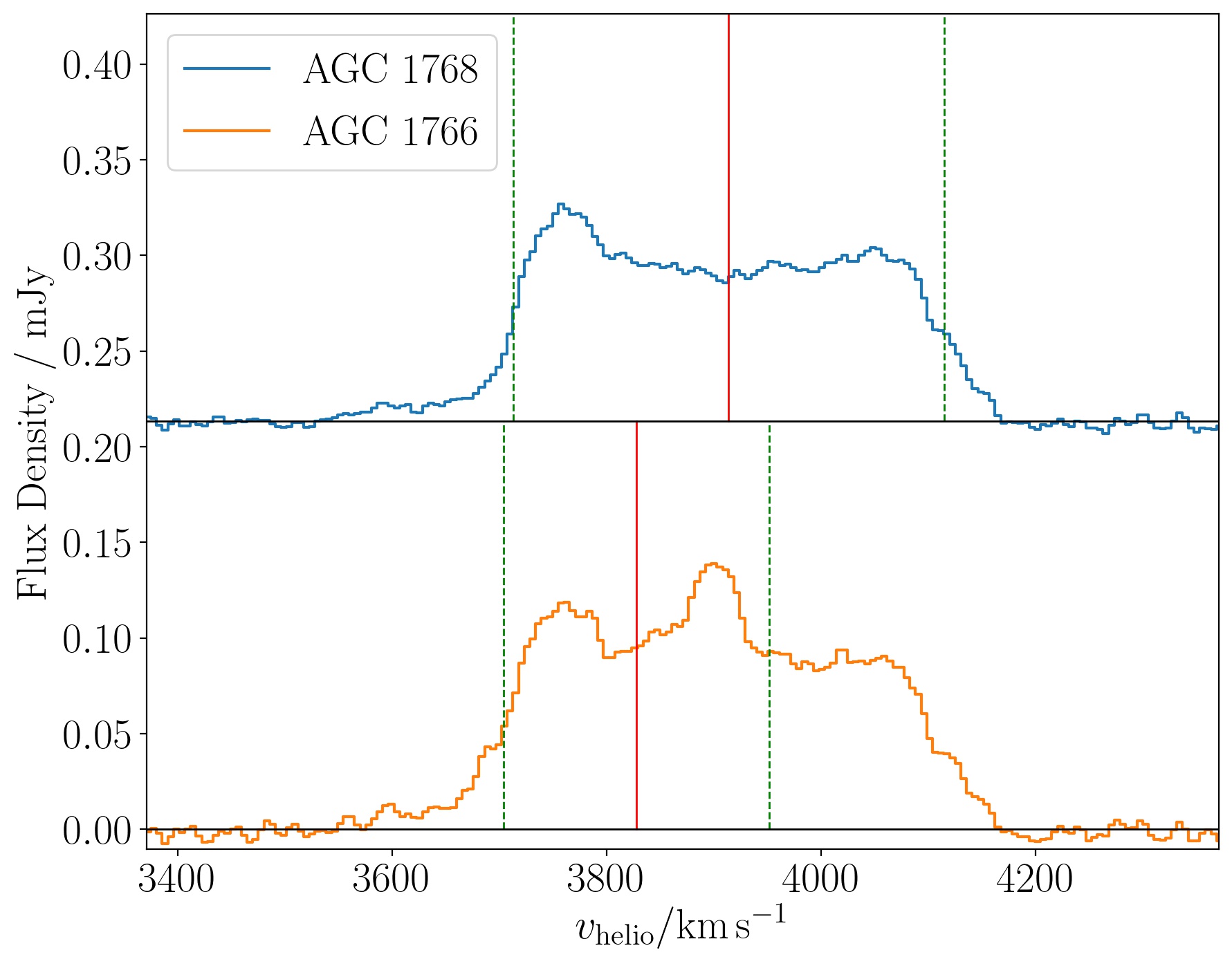}
\includegraphics[width=7cm]{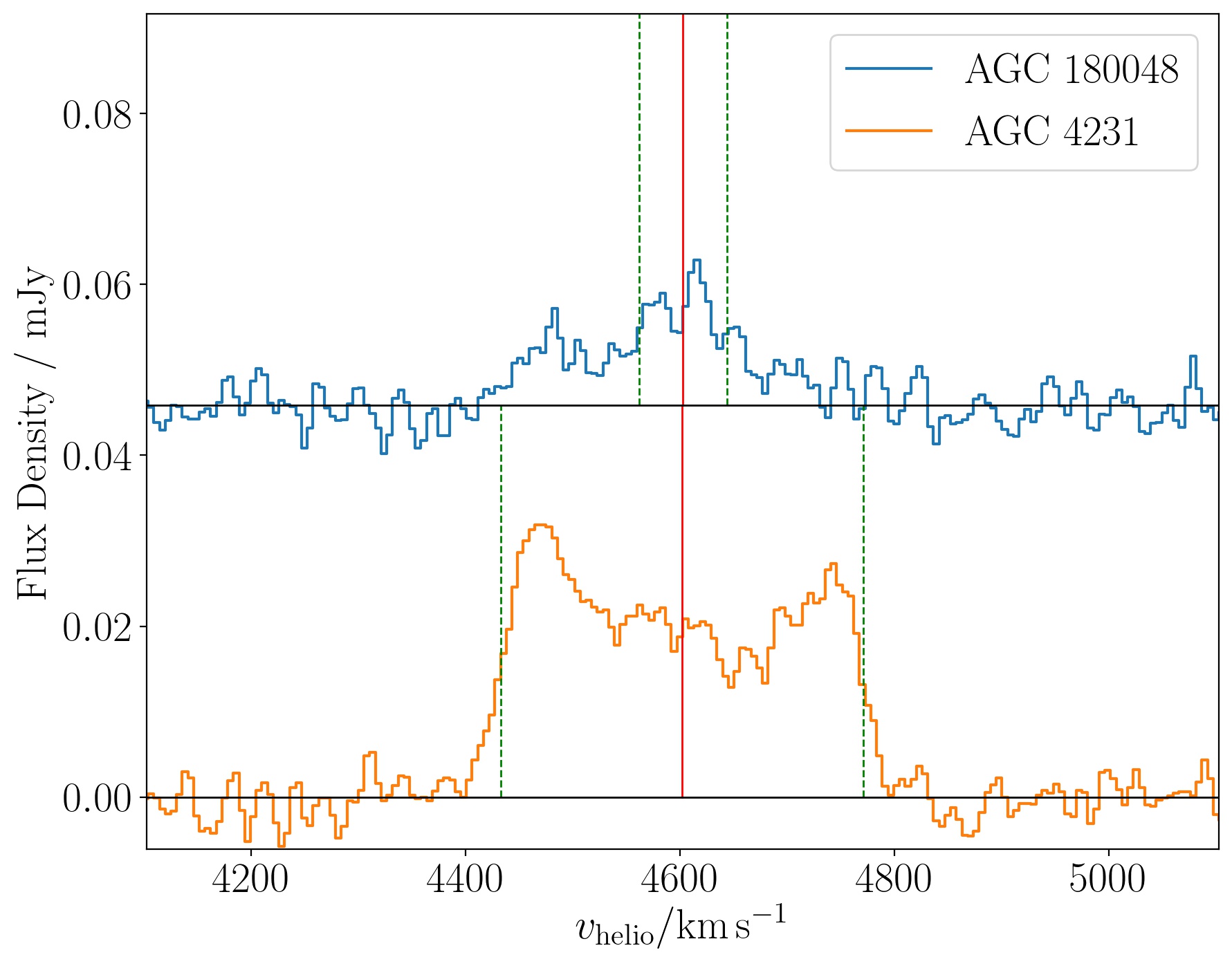}
\includegraphics[width=7cm]{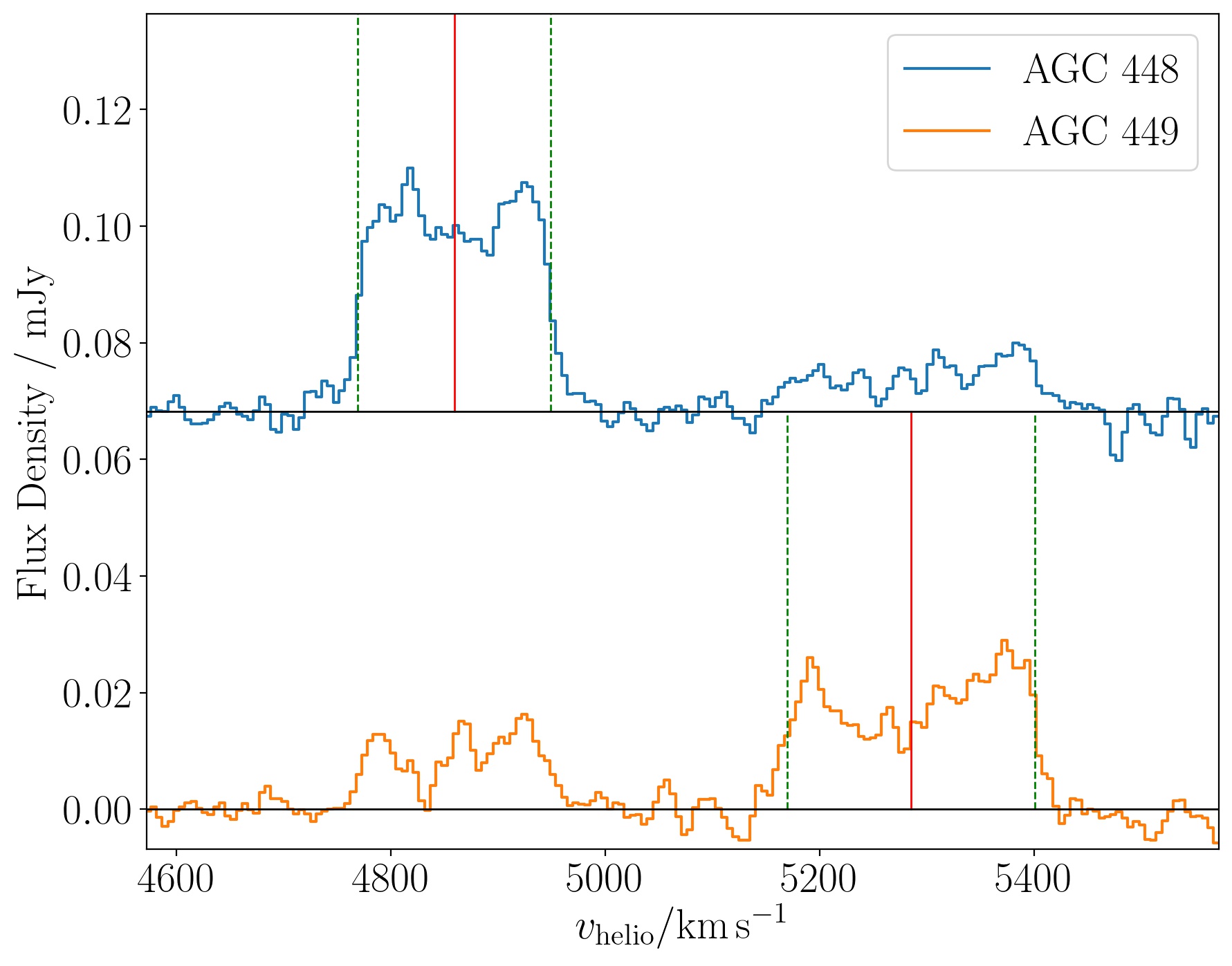}
\includegraphics[width=7cm]{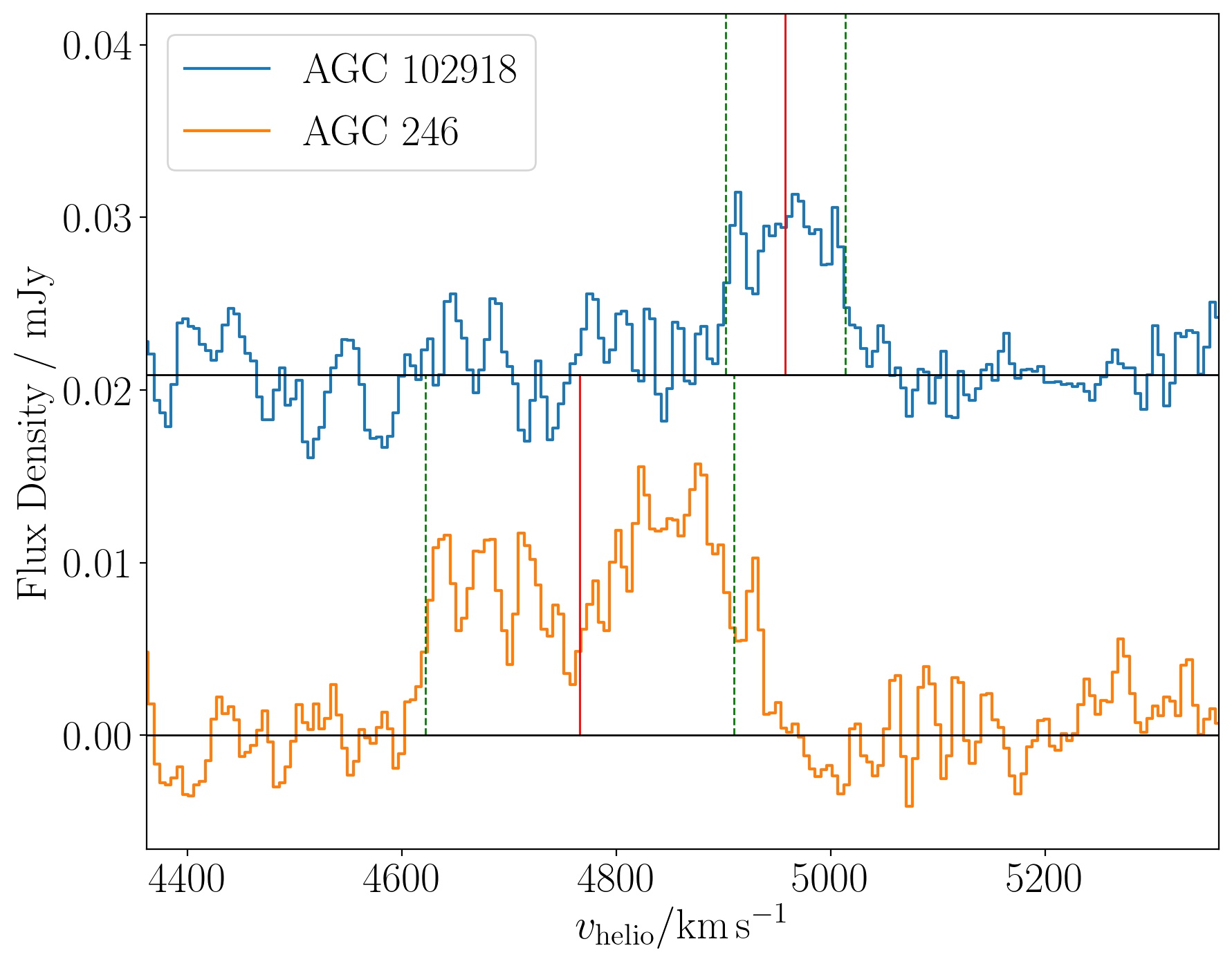}

\includegraphics[width=7cm]{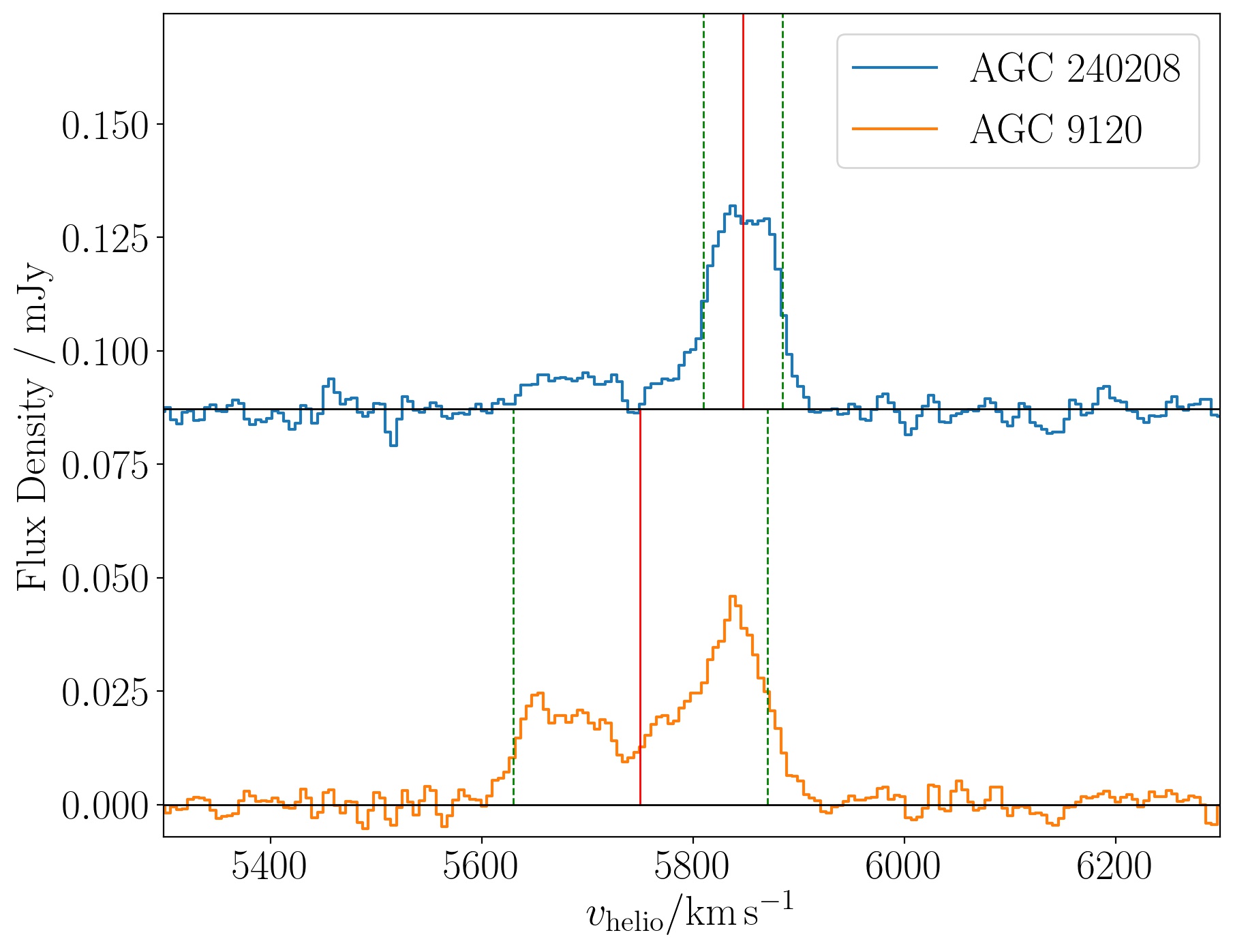}
\includegraphics[width=7cm]{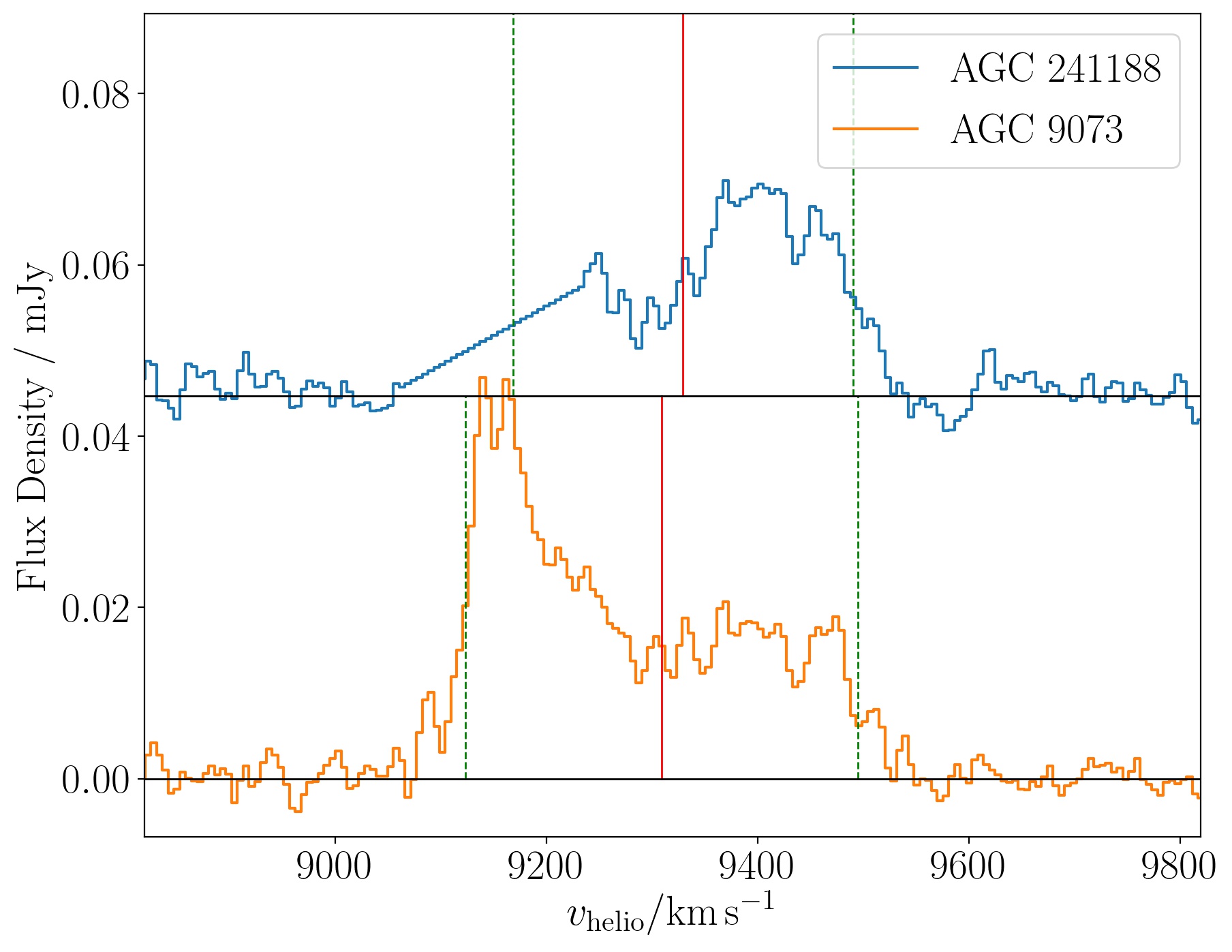}

\caption{Representative cases of HI blending in the pair sample.}
\label{HIblening}
\end{figure*}

We exclude a total of 49 pair members from our HI analysis on the basis of being in a blend too severe to render reliable HI data, as reported by the ALFALFA extraction team. A representative example of such a case is demonstrated by the pair AGC 1766/1768, where it is impossible to allocate which flux belongs to each galaxy due to their spectra completely overlapping in velocity space (see first plot in Figure \ref{HIblening}). In this instance both galaxies were removed from our analysis. The second plot is an example of a dwarf companion (AGC 180048) sitting on the HI flux pedestal of its parent (AGC 4231). We exclude AGC 180048 from our sample since it is impossible to accurately measure its flux, however keep AGC 4231. The remaining plots in Figure \ref{HIblening} are examples of pair members that were not excluded from our analysis despite being flagged as in some sort of blend by the ALFALFA team. In these instances the blend is noted as not severe, and the HI properties are regarded as reliable by the ALFALFA team. The pair members AGC 448/449, for example, are very close together on the sky and therefore noted as in a blend, but separated enough in velocity to prevent significant blending of their profiles. No significant blending was reported for the pair AGC 246/102918 since the sources are separated just enough on the plane of the sky. The pair AGC 240208/9120 is an example where there is definite blending between the sources, but it is does not appear to be severe. The blending here was noted, but both sources were kept. In the pair AGC 241188/9073, one side of the profile was extrapolated by the ALFALFA team to prevent significant blending. This was noted, but both sources were kept. These cases are extremely rare.

\bsp	
\label{lastpage}
\end{document}